\documentclass[10pt, a4paper]{article}

\usepackage{graphicx}
\usepackage{commath}
\usepackage{amsthm}
\usepackage{hyperref}
\usepackage{color}
\usepackage[table]{xcolor}
\usepackage{multirow}
\usepackage{booktabs}

\usepackage[english]{babel}
\usepackage{amssymb,amsmath,amsthm, amsfonts}

\RequirePackage[left=2cm,%
                right=2cm,%
                top=2.25cm,%
                bottom=2.25cm,%
                headheight=12pt,%
                letterpaper]{geometry}%

\theoremstyle{definition}\newtheorem{definition}{Definition}
\newcommand\Tstrut{\rule{0pt}{2.6ex}}

\title{Improving the pseudo-randomness properties of chaotic maps using deep-zoom}

\author{Jeaneth Machicao$^1$, Odemir Martinez Bruno$^{1}$}
\date{}
\begin{document}
\maketitle
\noindent{$^1$S\~{a}o Carlos Institute of Physics, University of S\~{a}o Paulo, S\~{a}o Carlos - SP, PO Box 369, 13560-970, Brazil.\\Scientific Computing Group - http://scg.ifsc.usp.br}

\begin{abstract}
A generalized method is proposed to compose new orbits from a given chaotic map.  The method provides an approach to examine discrete-time chaotic maps in a ``deep-zoom'' manner by using $k$-digits to the right from the decimal separator of a given point from the underlying chaotic map. Interesting phenomena have been identified. Rapid randomization was observed, i.e. chaotic patterns tend to become indistinguishable, when compared to the original orbits of the underlying chaotic map. Our results were presented using different graphical analyses (i.e., time-evolution, bifurcation diagram, Lyapunov exponent, Poincar\'{e} diagram and frequency distribution). Moreover, taking advantage of this randomization improvement, we propose a Pseudo-Random Number Generator (PRNG) based on the $k$-logistic map. The pseudo-random qualities of the proposed PRNG passed both tests successfully, i.e. DIEHARD and NIST, and were comparable with other traditional PRNGs such as the Mersenne Twister. The results suggest that simple maps such as the logistic map can be considered as good PRNG methods. 
\end{abstract}

\section{Introduction}

Random numbers play a significant role in various fields and in many demanding applications~\cite{KnuthBook,PressPRNG,MonteCarloMethod} ranging from statistical mechanics, decision theory, calculus, games and the gambling industry to areas with more crucial objectives such as cryptography and computer simulations. Consequently, in recent years, considerable attention has been paid to generating such numbers.

It is worth mentioning that methods for generating random numbers can be broadly divided into two main approaches: True Random Number Generators (TRNGs) and Pseudo-Random Number Generators (PRNGs). On the one hand, TRNGs rely on physical sources such as throwing a dice or flipping a coin to more sophisticated processes such as radius material decay~\cite{NuclearDecayRandom,NuclearDecayRNG}, thermal noise in resistors~\cite{thermalnoiseRNG}, atmospheric noise~\cite{noiseRNG,atmospherical-RNG}, lava lamps~\cite{LAVA-RNG} and quantum physics~\cite{Photon-QRNG,Optical-QRNG}. However, there are certain challenges in terms of implementation, including additional special equipment and circuitries (often requiring either cooling or high voltages), limited speed, restricted rates of production, e.g 16 Mbps~\cite{QRNG-mercado}, and degradation of the statistical properties when implemented at a hardware and software level~\cite{flipingcoin-RNG}.

On the other hand, PRNGs can generate pseudo-random numbers depending on deterministic sources by inputting an initial seed to a given algorithm, such as the Linear Congruential Generator (LCG)~\cite{KnuthBook,PressPRNG}, the Mersenne Twister~\cite{MersenneTwister}, nonlinear LCG~\cite{NonlinearLCG-PRNG}, Linear Feedback Shift Registers (LFSR)~\cite{LFSR-PRNG}, among other numerical algorithms. The main advantage of PRNGs is their determinism (repeatability of the pseudo-random sequences), which for certain purposes is a desirable characteristic specially considering cryptography and one-way functions (hashing algorithms). Although PRNGs are periodic, due to technological advances, many PRNGs have a long periodicity that leads to this problem being ignored in practice. Moreover, PRNGs are efficient in terms of producing many numbers in a short time, they  are inexpensive and are easy to implement. Many of these classical PRNGs, such as the old fashion RANDU, have already been cryptanalyzed~\cite{SadHistoryRandomness, Turingbook}. Therefore, considerable effort has been made to develop alternative PRNG proposals to fulfill the need of having ``good'' sources of pseudo-randomness.

Over the last 30 years, many scientists have observed an interesting relationship between chaos and pseudo-randomness~\cite{Kocarev2001,Patidar, Baptista, Wolfram-RULE30,Alvarez2006} given that chaos reaches desirable properties for cryptographic and cryptoanalysis~\cite{machicao2015modes} purposes. The distinct properties of chaos, such as diffusion and confusion can, in principle, ensure pseudo-random numbers of high-quality, mainly because of the sensitivity to initial conditions, ergodicity, mixing properties, complex numerical patterns, relatively simple equations and their determinism~\cite{Alvarez2006,ArroyoPHD,Patidar,Anderson2010}. Therefore, significant progress has been reported, for instance chaos-based PRNGs were proposed using chaotic systems~\cite{ Chaos-PRNG1,Chaos-PRNG2,Chaos-PRNG3,Chaos-PRNG4,Chaos-PRNG5,Chaos-PRNG6,Chaos-PRNG7,Chaos-PRNG8}, cellular automata~\cite{CA-PRNG2,MachicaoLifeLike,CA-PRNG3}, quantum chaotic maps~\cite{quantum-PRNG1, quantum-PRNG2,quantum-PRNG3} and even mixing the physical chaotic sources, e.g. chaotic semiconductor lasers~\cite{Uchida-laser-PRNG, Reidler-laser-PRNG2}.

It is worth remembering to the first chaos-based cryptosystem that appeared in the late 1990s proposed by Baptista~\cite{Baptista}, who explored the pseudo-randomness advantages from the logistic map as a PRNG, i.e. generating sequences of bits to  be used  later by the cipher ~\cite{Baptista}. However, the relationship between chaos and cryptography has been questioned, criticized and discouraged after Baptista's cryptosystem fault~\cite{AlvarezCriptonalise,Persohn2012}. Some researchers expressed certain doubts about  the pseudo-randomness properties of the logistic map such as (i) non-uniform probability distribution (U pattern), where a plateau distribution is expected~\cite{AlvarezCriptonalise}; ii) dependence on the control parameter, consequently leading to periodic windows~\cite{AlvarezCriptonalise}; iii) sufficiently large samples of ciphertext which can estimate the parameter~\cite{AlvarezCriptonalise}; and (v) trajectories of short cycle lengths where long periodicity is expected depending on machine limitations~\cite{Persohn2012}; (vi) degradation of digital chaotic system issues~\cite{DynamicalDegradation}. 

Though, the true potential from chaos is being dismissed since many researchers have been neglected the fact that chaos analytically rely on the infinitesimal depth of precision digits from their orbit points. For instance, taking into account the well-known Mandelbrot set~\cite{mandelbrotbook}, when it is displayed on a computer screen, it reveals interesting but quite limited patterns. Nevertheless, when this single pattern is magnified repeatedly, a huge number of complex patterns can be distinguished, and, effectively, is in these magnification views where legitimate chaos occurs. Thus, higher computing precision is required to explore the deep-zoom level of a chaotic system and consequently investigate the pseudo-randomness properties of chaotic systems.

In this paper, we proposed an approach to improve the pseudo-randomness properties of a PRNG based on chaotic maps. The main idea is to perform a deep-zoom exploration of the chaotic orbits in order to generate more chaotic orbits. In this article, we focus on the logistic map, as it is very well known in the chaos theory.
Specifically, each $x^{t}$ value from an orbit was explored by considering the $k$-right digits of precision, which we coined as the $k$-logistic map for the orbits derived from the underlying orbit $k=0$. Furthermore, the PRNG based on the $k$-logistic map was subject to a statistical randomness test in order to show the strong properties from the chaotic systems.

In the remainder of this paper, Section~\ref{Sec:preliminaries} presents a short background regarding round-off errors and dynamical degradation of chaotic systems. Section~\ref{Sec:proposal} outlines the mathematical definitions of the proposed approach to later introduce the $k$-logistic map. We divided the experiments into two parts. Firstly, Section~\ref{Sec:analysis} shows the overall dynamic properties of the $k$-logistic map. Secondly, Section~\ref{sec:prnganalysis} presents the pseudo-randomness analysis by means of the DIEHARD~\cite{DIEHARD} and NIST~\cite{NIST-PRNG} suites, and finally a comparison analysis is performed with another chaotic map and classical PRNGs. Section~\ref{sec:discussions} ends with discussions about the results and conclusions are drawn in Section~\ref{sec:conclusions}.

\section{Preliminaries}
\label{Sec:preliminaries}

\subsection{Influence of round-off errors and dynamical degradation of chaotic systems}

The pillar of the chaos theory states that the slightest change in the initial conditions of a chaotic dynamical system will eventually lead to an exponential separation from the original orbit~\cite{OttChaosBook}. Consequently, digits of precision are an important issue for numerical studies of chaotic systems. For instance, consider the logistic map given in Eq.~\ref{eq:logisticmap}
\begin{equation}
\label{eq:logisticmap}
  x_{t+1}= f(x_{t}) = \mu x_{t}(1-x_{t})\,,
\end{equation}
where $\mu \in [0,4]$, $x \in [0,1]$ and $t$ is the discrete time step~\cite{OttChaosBook}; and consider an orbit of the logistic map constituted analytically with parameter $\mu=4$. It can be observed how the number of digits of precision exponentially expand while the map iterates,

\noindent
$x_0=0.4$, \\
$x_1=0.96$,\\
$x_2=0.1536$,\\
$x_3=0.52002816$,\\
$x_4=0.9983954912280576$,\\
$x_5=0.00640773729417263956570612432896$,\\ 

Floating-point arithmetic is a very important topic in computer science. In many computations, the precision of the results are progressively degraded as a result of the ``round-off error''. A typical case of this issue is exemplified in Table~\ref{tab:exemplodigitsprecision}. 

\setlength{\tabcolsep}{3pt}
\begin{table}[h!]
\centering
\caption{Round-off applied to three orbits (columns) generated by the logistic map with the same initial conditions and parameters using $4,\ldots,6$ computing digits of precision.}
\label{tab:exemplodigitsprecision}
\begin{tabular}{clll}
& \multicolumn{3}{c}{computing digits of precision}   \\
$t$ & \multicolumn{1}{c}{4} & \multicolumn{1}{c}{5} & \multicolumn{1}{c}{6} \Tstrut \\
\hline \Tstrut 
0  & 0.4782 & 0.47820 & 0.478200 \\
1  & 0.9981 & 0.99810 & 0.998099 \\
2  & 0.0076 & 0.00759 & 0.007590 \\
3  & 0.0302 & 0.03013 & 0.030130 \\
\vdots  & &\multicolumn{1}{c}{\vdots} \\
100 &0.3997 & 0.68170 & 0.246600 \\
\end{tabular}
\end{table}

In this experiment round-offs to the $4,\ldots,6$-th computing digits of precision were used, while computing $t=100$ time steps in the logistic map using the same initial condition and parameters $\mu=4$. It can be observed that, in the long-term $x_{100}$, each of the orbits have departed away from the others orbits.

On the other hand, the so-called dynamical degradation~\cite{DynamicalDegradation} occurs when a chaotic system is done in finite computing precision with $L$-bit finite precision and fixed-point arithmetic is adopted. Thus, the initial conditions, control parameters and the orbit values are formulated as $a/2^{L}$ with $a = 0 \sim 2^L - 1$, and therefore there are only $2^L$ digital values to represent the chaotic orbits, therefore the period length of the orbits will not be larger than $2^L$~\cite{Li2001}. For this reason, it is not feasible to compute the orbit with the same precision as that obtained analytically. However, one way to circumvent these two problems is by using higher computing precision. Thus, in this manuscript we used the Apfloat~\footnote{\url{http://www.apfloat.org/}} library, which is a high performance arbitrary precision arithmetic used in all the experiments presented here.

\subsection{Number of iterations to reach the chaos regime}

The number of iterations required so that two orbits, initially slightly separated, are set apart considerably is also related to the round-off problem. For instance, we estimated the number of iterations $\tau$ needed to reach $|f^{\tau}(x) - f^{\tau}({x^\prime})| > 10^{-1}$ when given an initial condition $x_0$ and another initial perturbed condition $x^{\prime} =x_0 + \epsilon$ are separated apart by a distance of $\epsilon = 10^{-d}$ when we use parameter $\mu=4$, which is well known for reaching the chaotic regime. Here, the exponent $d$ represents the number of digits of precision $d = \lfloor \log |x_0 - x^{\prime}| \rfloor$. 

\begin{figure}[h!]
\centering
{\includegraphics[scale=0.5]{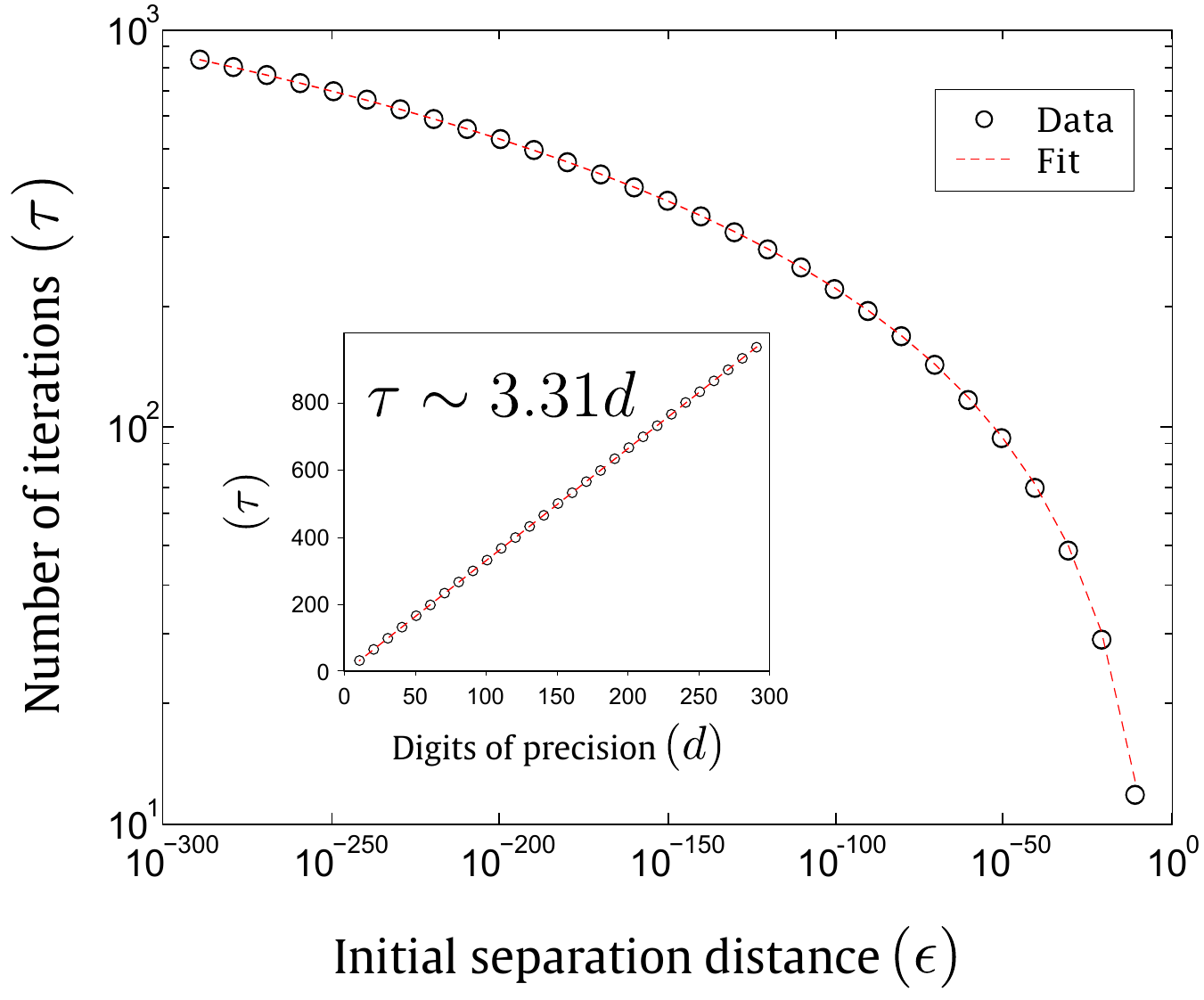}}
\caption{Log-log plot of the number of iterations $\tau$ needed to achieve $f^{\tau}(x) - f^{\tau}({x^\prime}) > 0.1$ when two orbits are separated apart at a distance $\epsilon$ (solid line). The inner plot shows the same results where the initial distance is shown in terms of the digits of precision $d$. The function $f(x)$ was related to the logistic map (Eq.~\ref{eq:logisticmap}). The goodness statistic of the fit curve $\tau \sim 3.311 d$ (dotted red line) are $R^2=0.99$ and ${\chi}^2= 56.2$.}
\label{fig:precisiondelta}
\end{figure}

Fig.~\ref{fig:precisiondelta} depicts a log-log plot of data obtained from several random initial conditions and its separation distances $\epsilon =\{10^{-10}, 10^{-20}, \ldots, 10^{-290}\}$, while the inner plot shows the same results but the initial separation distance is configured in terms of the digits of precision $d=\{10, 20, \ldots, 290\}$. Regarding this graph, we found a linear approximation given by $\tau \sim 3.311 d$, which suggests that the number of iterations $\tau$ needed so that two initial orbits, initially slightly separated, move away from each other should be at least three times the number of $d$ digits of precision being used.

\begin{figure*}
{\includegraphics[width=0.99\textwidth,height=0.3\textwidth]{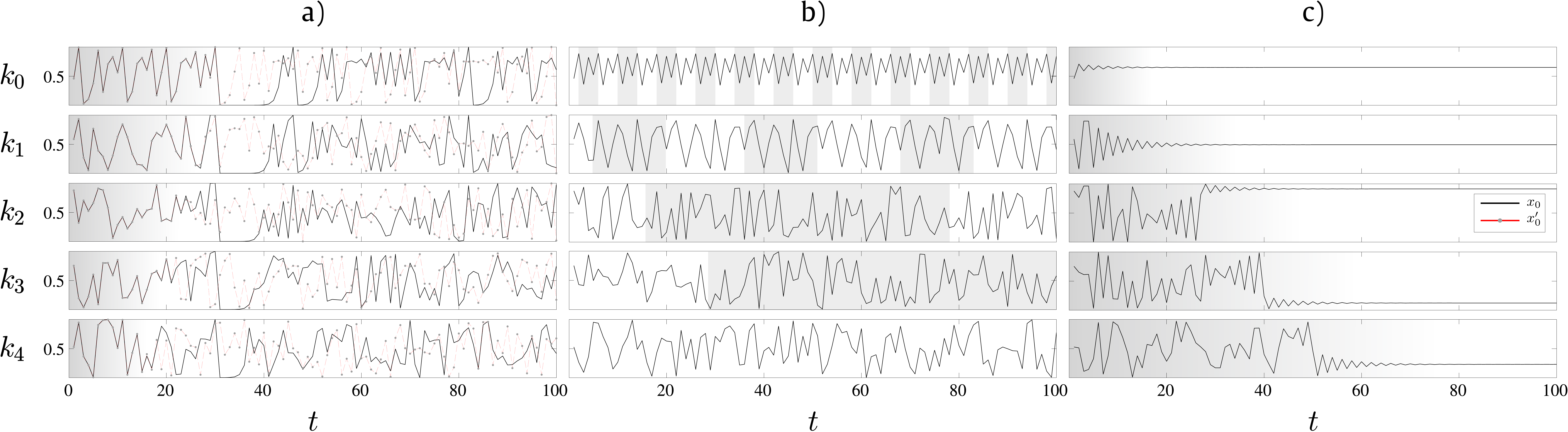}}
\caption{Time-evolution of two orbits with close initial conditions $x_0=0.4587525281$ (solid line) and $x_0^{\prime}= 0.4587525282$ (dotted line) using the parameters (a) $\mu=4.0$ (b) $\mu=3.57$ and (c) $\mu=2.85$, over $t=100$ iterations, considering from $k=0$ to $k=4$ decimal digits (top to down). The light gray watermark represents a) the number of iterations for the two orbits which diverge from each other, b) the length of the periods and c) the number of iterations until the orbit converges.}
\label{fig:timeevolution}
\end{figure*}

\section{k-right digits of the Logistic map}
\label{Sec:proposal}

A one-dimensional discrete dynamical system (DDS) can be described by a recurrence equation $x_{t+1}= f(\mu, x_t)$ such that $f:M\mapsto M$ is a map where $x_t$ at a particular time $t \in \mathbb{N}_0$ represents all the information characterizing a system given an initial state $x_0$ in the phase-space $M \subset \mathbb{R}$, which depends on the control parameter $U \subset \mathbb{R}$. Therefore, an orbit $\mathcal{O}(\mu, x_0)=\{x_0, x_1,\ldots, x_t\}$ is a collection of points related by the evolution function of the DDS. For instance, the logistic map (Eq.~\ref{eq:logisticmap}) is formalized as $f(\mu,x_{t+1})=\mu x_{t}(1-x_{t})$ such that $x_t\in M=[0,1]$ and $\mu \in U=[0,4]$. 

\begin{definition}
Let $\mathcal{O}^{k}(\mu,x_0)= \{x_0^k, x_1^k,\ldots, x_t^k\}$ the resulting orbit derived from an already observed orbit $\mathcal{O}(\mu,x_0)$, or simply $\mathcal{O}^{k=0}$. The new value $x_t^k$ is created by retaining $k$-decimals to the right of the decimal separator corresponding to a given point $x_t \in \mathcal{O}^{k=0}$. Thus, Eq.~\ref{eq:klogistic} is the generalized approach to analyze the $k$-right digits of precision of a one-dimensional DDS
\begin{equation}
	x_t^k= \phi_{k,L}(x_t) = \frac{\lfloor x_t 10^{k+L}\rfloor}{10^L} - \lfloor x_t10^k \rfloor\,,
	\label{eq:klogistic}
\end{equation}
\noindent where $L$ is the length of digits of precision of the new point $x_t^k$, and $\lfloor$ $\rfloor$ stands for the floor function.
\end{definition}
Clearly, some constraints must be taken into account. 
\begin{itemize}
\item $k \leq L$;
\item $k+L \leq \text{length}\{x_t^{k=0}\}-2$, otherwise the resulting value is padded with zeros;
\item the transient time, i.e the number of iterations before the systems settle down must be at least three times the number of digits of precision being used.
\end{itemize}
Moreover, the number of digits precision $L$ should not be confused with the computing precision of the underlying configuration, which in practice should use the highest precision as possible.

\begin{definition}
\label{def:k-logisticmap}
The resulting orbit $\mathcal{O}^k(\mu,x_0)$, hereafter $k$-logistic map, is a straightforward function to generate orbits derived from an underlying orbit $\mathcal{O}^{k=0}(\mu,x_0)$, such that the resulting orbit still maintains its parameters $\mu \in [0,4]$, $x_t^k$ in the unitary interval $[0,1]$. 
For example, given two points $x_{t}=0.268932$ and $x_{t+1}=0.786430317504$ of an orbit $\mathcal{O}^{k=0}$. Taking a step to the right from the three first digits to the right of the decimal point, we obtain for example $\phi_{k=1,L=3}(x_t)= 0.689$ while $\phi_{k=1,L=3}(x_{t+1})= 0.864$. Another example, $\phi_{k=3,L=4}(x_t)= 0.9320$ while $\phi_{k=3,L=4}(x_{t+1})=0.4303$, and so on.
\end{definition}

\begin{definition} 
\label{def:k-LE}
Furthermore, some formal definitions of the Lyapunov exponent for one-dimensional maps are recalled below. Typically, the LE can be assessed analytically by using Eq.~\ref{eq:normalLE}. 
\begin{equation}
\label{eq:normalLE}
\lambda(\mu,x_0) = \lim_{T\to\infty}\frac{1}{T}\sum_{t=0}^{T-1} \ln\abs{f(\mu,x)^\prime(x_t)}\,.
\end{equation}
Therefore, we can approximate the LE for the new constituting orbit $\mathcal{O}^k$ by using Wolf's method to seek Lyapunov exponents in times series~\cite{Wolf-LE}. Given two orbits in the phase space with initial separation $d_0 = \abs{\phi_{k}(f(x_0+\epsilon)) - \phi_{k}(f(x_0))}$ diverge at a rate given by $\epsilon e^{t \lambda (x_0)}$ for long term $T$. The LE for one-dimensional maps $\lambda(\mu, x_t^k)$, such as the $k$-logistic map is given by Eq.~\ref{eq:klyapunov}, as follows
\begin{equation}
\label{eq:klyapunov}
\lambda(\mu,x_t^k) = \frac{1}{T \Delta t}\sum^{T}_{t=1}\ln \biggl(\frac{d_t}{d_0} \biggr)\,,
\end{equation}
where $\Delta t$ is the number of time steps in the fiduciary trajectory, and $T$ is the number of times through the loop inside Eq.~\ref{eq:klyapunov}, and 
$d_t = \abs{\phi_{k}(f^{\Delta t}(x_0+\epsilon)) - \phi_{k}(f^{\Delta t}(x_0))}$. For more suitable numerical computations, we discretize the time into the interval of length $\Delta t$, such that the total number of time-steps is $T\Delta t$.
\end{definition}

\section{Analysis of k-logistic map}
\label{Sec:analysis}

In this section, we analyzed the overall dynamic properties of the orbits generated by the $k$-logistic map (Eq.~\ref{eq:klogistic}) by means of several plots including time-evolution (Section~\ref{sec:timeevolution}), the bifurcation diagram and skeleton bifurcation diagram (Section~\ref{sec:biff}), statistical analysis with Lyapunov exponents (Section~\ref{sec:lyapunov}), phase diagrams (Section \ref{sec:phasediagram}) and frequency distribution plots (Section~\ref{sec:freqdistrib}). Moreover, for all of these experiments, we employed $500$ digits of precision by using the Apfloat library, so that we could expand them to generate new values as $k$ increases. Additionally, hereinafter the terms $k_i$ and $k=i$ are used indistinctly.

\begin{figure*}[!ht]
\centering
{\includegraphics[scale=0.28]{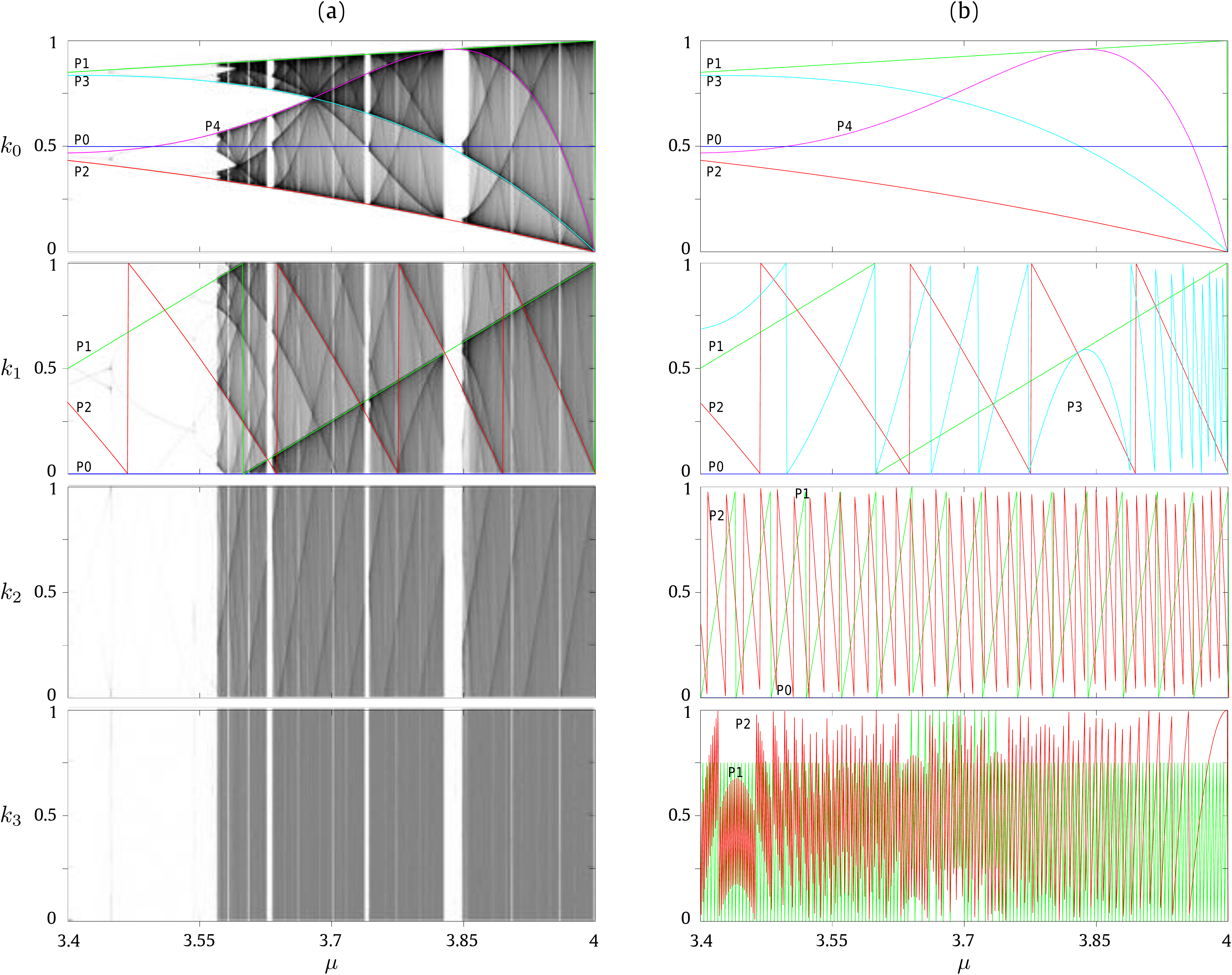}}
\caption{a) Bifurcation diagram of the $k_0$, $k_1$, $k_2$ and $k_3$-logistic map (Eq.~\ref{eq:klogistic}). The horizontal axis depicts $\mu \in [3.4,4]$ (steps of $0.001$). The vertical axis shows the possible long-term values of the corresponding $k$-logistic map, starting from the same initial condition, calculated over $10^5$ iterations (first $200$-th are the transient). b) The superimposed curves $P_n(\mu)$ (Eq.~\ref{eq:skeletonk}) correspond to the skeleton bifurcation (colored lines).}
\label{fig:biffkmapalogistico}
\end{figure*}

\subsection{Time-evolution}
\label{sec:timeevolution}

Fig.~\ref{fig:timeevolution} depicts the time-evolution of the $k$-logistic map corresponding to the orbits from $k_0$ (original) to $k_4$ during $t=100$ iterations. Here we used specific parameters $\mu=4$, $\mu=3.57$ and $\mu=2.85$ that lead to the three well-known Lyapunov stability behaviors, i.e. chaotic, periodic and stable respectively. The plots in Fig.~\ref{fig:timeevolution}-a) show two nearby close orbits with initial conditions $x_0$ and $x_t^{\prime}$, using the parameter $\mu=4$ that leads to the chaotic regime. It can be observed that the transient time (before both curves diverge) is reduced approximately from 31 ($k_0$) to 17 ($k_4$). Fig.~\ref{fig:timeevolution}-b) depicts the plots on the periodic regime. It can be observed that there is a periodic pattern ($k_0$) that is repeated almost 24 times across the $t=100$ iterations, while the periodic sequence on $k_1$ is repeated 6 times, in $k_2$ it is repeated almost 2 times, in $k_3$ and $k_4$ the period length is bigger than $t=100$ iterations and cannot be observed in this plot. Finally, observing the steady regime in Fig.~\ref{fig:timeevolution}-c), the number of iterations before reaching a convergence to a fixed point is increased successively around 11 ($k_0$) to 71 ($k_4$).

Regarding these graphs, we observed two things: that the curves are stretched and contracted, depending on the control parameter $\mu$ while $k$ is increased. Notwithstanding, the corresponding LE stability of the original curve ($k_0$) is maintained as $k$ increases, for instance in the chaotic region, the transient time is decreased; in the periodic region, the cycles have bigger lengths and the number of iterations on the stable regime before the convergence is increased.

\subsection{Bifurcation Diagram}
\label{sec:biff}

Fig.~\ref{fig:biffkmapalogistico}-a), from top to down, shows the well-known bifurcation diagram for the original logistic map ($k=0$) with control parameter $\mu$ in $[3.4, 4]$, and the corresponding bifurcation diagram for the $k_1$, $k_2$ and $k_3$-logistic map. We also computed the first skeleton bifurcation curves~\cite{HaoBailinBook} (see Fig.~\ref{fig:biffkmapalogistico}-b)) from the critical point $x_c= 0.5$ as a parameter of the function $P_n(\mu)$ given in Eq.~\ref{eq:skeleton}, and described in detail in Appendix~\ref{sec:AppendixA}, which depicts the colored lines superimposed on the bifurcation diagram.
\begin{equation}
\label{eq:skeleton}
P_{n+1}(\mu) = \mu P_n(1-P_n(\mu)) \,,
\end{equation}
where $P_0(\mu)= 0.5$. Moreover, we also plot the first skeleton bifurcation curves for the $k_1$ to $k_3$-logistic map by extending the former equation and extracting the $k$-decimal digits using Eq.~\ref{eq:klogistic} as follows
\begin{equation}
\label{eq:skeletonk}
P^k_{n+1}(\mu) = \phi_{k,L}(P_n(\mu)).
\end{equation}

It can be observed that the interspersed chaotic patterns nearly at the regions $\mu \in [3.57, 3.828]$ and $\mu \in [3.86, 4]$ are in a zigzag pattern ($k_1$) and start to gradually vanish ($k_2$) until they seem random ($k_3$). However, the ``islands of stability'', which can be easily visualized as the white strips of the bifurcation diagram (for instance intervals $3.4$ to $3.57$ and $3.828$ to $3.86$), remain constant for the whole $k$-logistic map. Regarding the period-doubling bifurcation points, it can be observed that their patterns become more filled as $k$ increases. In some cases, the periodic regions, where the bifurcations intervals occur, are multiplied and the points are spread into the plot. It should be mentioned that these phenomena were also observed in the time-evolution plot (see Fig.~\ref{fig:timeevolution}). The zigzag behavior can also be seen in the skeleton bifurcation (Fig.~\ref{fig:biffkmapalogistico}-b), since their curves are in a zigzag pattern ($k_1$ and $k_2$) and are saturated rapidly while $k$ increases ($k_3$).

\subsection{Lyapunov stability}
\label{sec:lyapunov}

\begin{figure}[!hbtp]
\centering
{\includegraphics[scale=0.44]{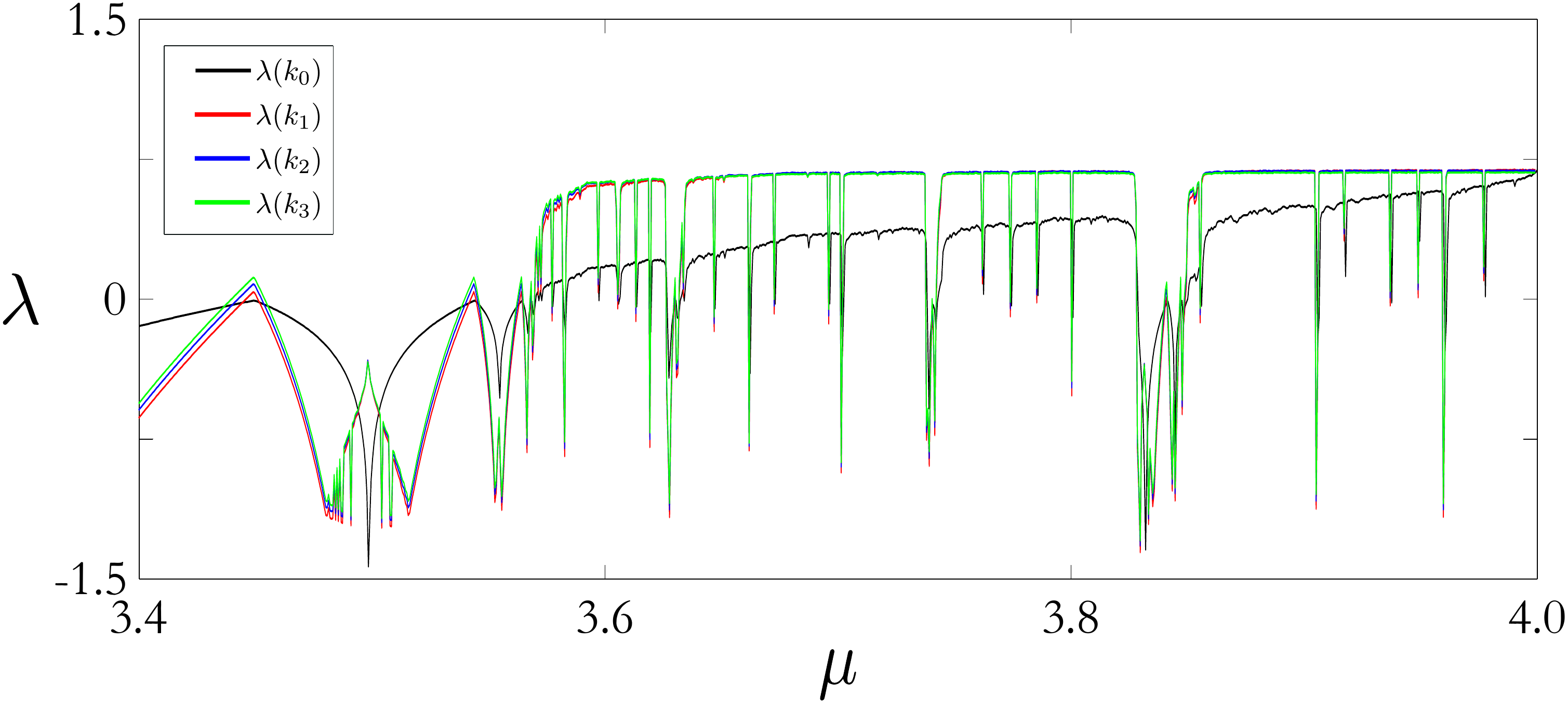}}
\caption{Lyapunov exponents of the $k$-logistic map for the $\lambda( k_0)$, $\lambda(k_1)$, $\lambda(k_2)$ and $\lambda(k_3)$. The horizontal axis depicts the LE in the region $\mu \in [3.4, 4]$ with steps of $0.0006$. The LEs were computed using $T=1000$ iterations and $\delta t=50$ starting from $d_0= 10^{-16}$.}
\label{fig:biffkmapaLE}
\end{figure}

Fig.~\ref{fig:biffkmapaLE} shows the approximate Lyapunov exponent (Eq.~\ref{eq:klyapunov}) of the $k$-logistic map orbits in the region $\mu \in [3.4, 4]$, which depicts the three types of dynamical behaviors, i.e. chaotic ($\lambda \geq 0$), periodic ($\lambda \approx 0$) and stable ($\lambda < 0$). 

In the chaotic region $\mu \in [3.57, 3.828]$ and $\mu \in [3.86, 4]$, the positive LE curves of $k_1$, $k_2$ and $k_3$ can be observed that are found almost below the LE $\lambda(\mu,k_0)$. In fact, there was an increase in the LE that complies with the upper bound found analytically at $\lambda = \ln 2\approx 0.693$ for all of the LE curves $\lambda(\mu,x^k)$ obtained by using Eq.~\ref{eq:klyapunov}. On the other hand, in the non-chaotic regions, i.e. in the stable and periodic stability regions which includes the small gaps of the chaotic region (so-called island of stability), a corresponding negative and zero LE curves can be observed for all of the $\lambda(\mu,x^k)$ curves. These results indicate that the orbits of the $k$-logistic are mandatory asymptotically stable or periodic in agreement with the behavior of the original orbit ($k_0$) as was previously shown in the bifurcation diagram and the time-evolution plot (see Fig.~\ref{fig:timeevolution}-\ref{fig:biffkmapalogistico}).

\subsection{Phase diagram}
\label{sec:phasediagram}

\begin{figure}[!ht]
\centering
{\includegraphics[width=8cm,height=15.8cm]{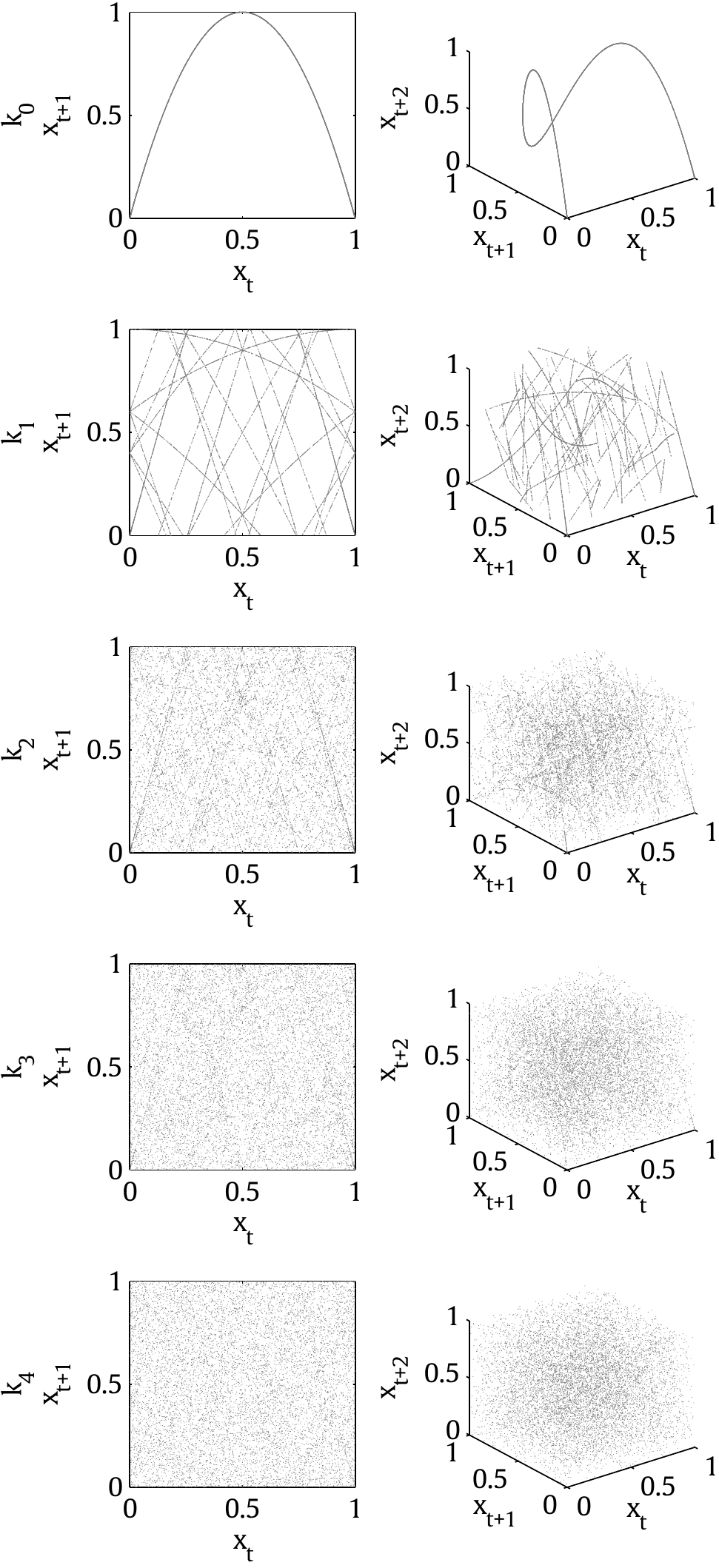}}
\caption{Phase diagrams for the $k_0$, $k_1$, $k_2$, $k_3$ and $k_{4}$-logistic map using $\mu = 4$. Two- and three-dimensional diagrams are shown in the left and right columns, respectively. The horizontal and vertical axes show the phase space of $x^k_t$ against $x^k_{t+1}$. Each orbit contains $10^4$ points starting from random initial conditions, where the first 200 iterations were discarded (transient time).}
\label{fig:poincareLogistic}
\end{figure}

Another way to visualize the pseudo-randomness properties of the $k$-logistic map is by means of the phase diagram, or Poincar\'{e} diagram, which maps the value of $x^k_{t+1}$ against $x^k_{t}$ of a given orbit, or even three-dimensional by mapping $x^k_{t+2}$, $x^k_{t+1}$ against $x^k_t$. For instance, Fig.~\ref{fig:poincareLogistic} depicts the phase diagrams of the $k$-logistic map for four different orbits $\mathcal{O}^{k}$ using $\mu=4$ as a control parameter.

From top to down, the $k_0$ shows the classical inverted parabola with a maximum value at $x=0.5$, whose pattern remains in the three-dimensional sketch. The row below shows the $k_1$ phase diagrams, were it can be observed that the former parabola was transformed into a zigzag pattern. The zigzag patterns are progressively being spread and vanish into a random plot as can be observed from $k_2$ and $k_4$. In fact, it can be observed that the phase space is filled, i.e. the $k$-logistic map produces almost all the possible values, while $k$ increases.

\subsection{Frequency distributions}

\label{sec:freqdistrib}

\begin{figure}[!ht]
\centering
{\includegraphics[scale=0.27]{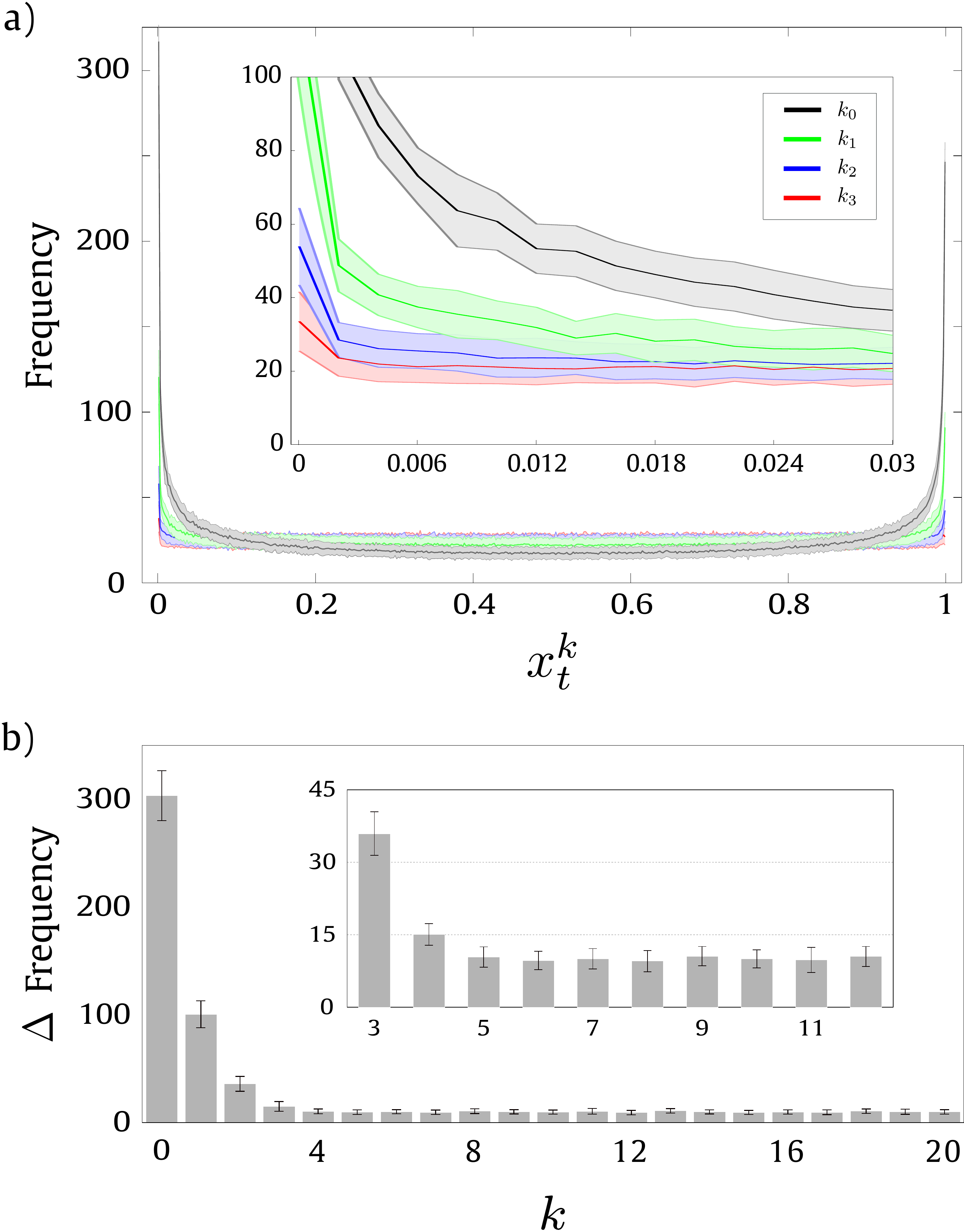}}
\caption{a) Frequency distribution curves for the $k_0$, $k_1$, $k_2$ and $k_3$-logistic map using $L=3$ and $\mu=4$. The horizontal axis shows the $x \in [0, 1]$ (500 bins) and the vertical axis shows the frequency of the $10^4$ values discarding the first $10^3$ transient values. The curves represent the mean and standard deviation (shaded error bar) for sequences generated over 100 random initial conditions. The inset plot depicts a zoom on the windows $x \in [0, 0.03]$. b) Deviation between the highest and lowest value ($\Delta$) obtained by the $k_0$, $k_1$,$\ldots$, $k_{20}$-logistic map frequency distribution curve. The inner plot shows a magnification of the $\Delta$ frequency plot in the region $k_3$ to $k_{11}$.}
\label{fig:histogramas}
\end{figure}

Fig.~\ref{fig:histogramas}-a) depicts the frequency distribution curves for orbits $k_0, k_1,k_2,$ and $k_3$ using parameter $\mu=4$. Concerning the $k_0$ curve (gray), it depicts the well-known U pattern that follows the invariant probability density $\displaystyle \rho(x) = 1/\pi\sqrt{1 - x^2}$ of the original logistic map~\cite{OttChaosBook}. 
Thus, the presence of higher frequencies at the boundary regions $x \in [0, 0.1]$ and $x \in [0.9, 1]$ can be observed more than in the central region.
%

However, at the inset plot of Fig.~\ref{fig:histogramas}-a), we noticed that curves $k_1,k_2$ and $k_3$ (colored) show a rapid decrease in the U pattern leading to a more uniform distribution as $k$ is increased. The reader should note that a plateau distribution for the purposes of PRNG should be expected. Moreover, this homogenization course is depicted in Fig.~\ref{fig:histogramas}-b), which shows the difference $\Delta$ between the highest and lowest frequency obtained for the $k_0$, $k_1$,$\ldots$, $k_{20}$-logistic map, which is in agreement with the randomization course illustrated in the phase diagram in Fig.~\ref{fig:poincareLogistic}.

\section{PRNG based on k-logistic map}
\label{sec:prnganalysis}

Before we introduce the PRNG based on the $k$-logistic map, first we must constrain Definition~\ref{def:k-logisticmap} given in Section~\ref{Sec:proposal}. Therefore, the initial condition $x_0$ must be limited to $]0,1[$. Additionally, the control parameter $\mu$ must be limited to the chaotic windows, e.g. $\mu \in [3.57, 3.828]$ and $\mu \in [3.86, 4[$. These limitations must be considered in order to avoid, for instance superstable points $f(\mu,0)=f(\mu,1)=0$ and critical point $f(4, 0.5)=1$, that lead to trivial values, and consequently, these are counterproductive for the generation of pseudo-random numbers. Yet, within the framework of PRNG purposes, we limited the PRNG based on the $k$-logistic map by using the control parameter $\mu \rightarrow 4$ as a prototype for the chaotic regime, as it represents the maximum LE (Eq.~\ref{eq:normalLE}), which  was conjectured by Collet\&Eckman\cite{Collet-Eckman2009}.

Therefore, in order to generate pseudo-random numbers using the $k$-logistic map, we simply discretized each $x_t$ value in order to obtain a 32-bit integer as follows
\begin{equation}
\label{eq:discretize}
\lfloor x_t^k (2^{32}-1)\rfloor 
\end{equation}
Thus, in order to evaluate the inherent randomness properties of the chaotic dynamics of the $k$-logistic map, we subjected orbits $k_0$, $k_1,\ldots, k_9$ to two well-known pseudo-randomness suites namely Diehard~\cite{DIEHARD} and NIST~\cite{NIST-PRNG}. Both suites provide different tests based on hypotheses under $p$-values, where the null hypothesis $H_0$ determines whether or not a specific binary sequence is accepted or rejected as a random, otherwise it is considered as a failure. These battery tests are configured to read source files as a stream of bits, and therefore can evaluate the randomness of the stream.

\subsection{Diehard battery}
\label{sec:diehardlogistic}

The Diehard test suite includes eighteen statistical tests (see more details in Ref.~\cite{DIEHARD}). According to the Diehard documentation, the binary sequences should be up to 10MBytes~\cite{DIEHARD}. Therefore, we generated 100 files (each containing 12.5MBytes) for each orbit $\mathcal{O}^{k}(\mu,x_0)$, considering the chaotic region under $\mu\rightarrow 4$. Each sample file was generated from a randomly chosen seed $x_0$, and each value $x_t^k$ was discretized as a 32-bit integer (Eq.~\ref{eq:discretize}) due to the Diehard restrictions~\cite{DIEHARD}. 
\newcommand{\pp}{ \cellcolor[HTML]{C0C0C0}} %
\begin{table}[!ht]
\caption{Average number of files that passed testes DIEHARD using $k$-logistic map PRNG (Eq.~\ref{eq:discretize}) from 100 files samples. Severe failed tests are shown in gray. All tests are passed at the interval $0.0001<\text{p-value}<0.9999$.}
\label{tab:diehardlogisticmap}
\begin{tabular}{ll}
\end{tabular}
\small
\centering
\begin{tabular}{lrrrrrrrrrr}
\toprule
\textbf{Diehard tests}  & $k_0$ &$k_1$ & $k_2$ & $k_3$ & $k_4$ & $k_5$ & $k_6$ & $k_7$ & $k_8$ & $k_9$ \Tstrut\\
\hline \Tstrut
BirthdaySpacings [KS] & 100 & 100 & 100 & 100 & 100 & 100 & 100 & 100 & 100 & 100 \\
OverlappingPermutations & 99 & 97 & 98 & 95 & 98 & 96 & 98 & 98 & 99 & 100 \\
Ranks31x31 matrices & 100 & 100 & 100 & 100 & 100 & 100 & 100 & 100 & 100 & 100 \\
Ranks32x32 matrices & 100 & 100 & 100 & 100 & 100 & 100 & 100 & 100 & 100 & 100 \\
Ranks6x8 matrices [KS] & \pp0 & \pp0 & \pp25 & 99 & 100 & 100 & 100 & 100 & 100 & 100 \\
Monkey20bitsWords [KS] & \pp0 & 99 & 100 & 100 & 100 & 100 & 100 & 100 & 100 & 100 \\
OPSO [KS] & 98 & 99 & 100 & 100 & 100 & 100 & 100 & 100 & 100 & 100 \\
OQSO [KS] & 98 & 100 & 100 & 100 & 100 & 100 & 100 & 100 & 100 & 100 \\
DNA [KS] & 100 & 100 & 100 & 100 & 100 & 100 & 100 & 100 & 100 & 100 \\
Count1sStream & \pp0 & \pp0 & \pp0 & 98 & 100 & 100 & 100 & 100 & 100 & 100 \\
Count1sSpecific [KS] & \pp0 & \pp0 & \pp0 & \pp0 & 94 & 100 & 100 & 100 & 100 & 100 \\
ParkingLot [KS] & 100 & 100 & 100 & 100 & 100 & 100 & 100 & 100 & 100 & 100 \\
MinimumDistance [KS] & 96 & 100 & 100 & 100 & 100 & 100 & 100 & 100 & 99 & 100 \\
RandomSpheres [KS] & 100 & 100 & 100 & 100 & 100 & 100 & 100 & 100 & 100 & 100 \\
Squeeze [KS] & 100 & 100 & 100 & 100 & 100 & 100 & 100 & 100 & 100 & 100 \\
OverlappingSums [KS] & 100 & 100 & 100 & 100 & 100 & 100 & 100 & 100 & 100 & 100 \\
Runs (up) & 100 & 100 & 100 & 100 & 100 & 100 & 100 & 100 & 100 & 100 \\
Runs (down) & 100 & 100 & 100 & 100 & 100 & 100 & 100 & 100 & 100 & 100 \\
Craps (wins) & 100 & 100 & 100 & 100 & 100 & 100 & 100 & 100 & 100 & 100 \\
Craps (throws/game) & 100 & 100 & 100 & 100 & 100 & 100 & 100 & 100 & 100 & 100\\
\bottomrule
\end{tabular}
\end{table}

\setlength{\tabcolsep}{2pt}
\newcommand{\p}{ \cellcolor[HTML]{C0C0C0}} 
\newcommand{\tabitem}{~~\llap{\textbullet}~~}
\begin{table}[!ht]
\caption{Number of files that passed the NIST test suits~\cite{NIST-PRNG} for $k$-logistic map. Failed tests are shown in gray. All tests are passed at the $\alpha=0.01$ significance level.}
\label{tab:nistLogistic}
\fontsize{9}{9}\selectfont
\centering


\begin{tabular}{lrrrrrrrrrr}
\toprule
NIST tests &$k_0$ &$k_1$ &$k_2$ &$k_3$ &$k_4$ &$k_5$ &$k_6$ &$k_7$ &$k_8$ &$k_9$ \\
\hline
\Tstrut
Frequency & 98 & 99 & 99 & 99 & 99 & 99 & 99 & 99 & 99 & 99 \\
\Tstrut
BlockFrequency $(m=128)$ & \p 0 & \p 1 & 66 & 95 & 98 & 98 & 99 & 100 & 99 & 99 \Tstrut\\
\hline  
\multicolumn{11}{l}{CumulativeSums}  \Tstrut\\
\hspace{1cm}Forward sums & 97 & 98 & 99 & 99 & 98 & 99 & 99 & 99 & 99 & 99 \\
\hspace{1cm}Reverse sums & 97 & 99 & 99 & 99 & 99 & 99 & 99 & 99 & 99 & 99 \\
\hline 
\Tstrut
Runs & \p 0 & \p 0 & \p 14 & 91 & 98 & 99 & 99 & 99 & 99 & 100 \\
\Tstrut
LongestRun & \p 0 & \p 0 & \p 15 &  89 & 98 & 99 & 98 & 100 & 99 & 99\\
\hline
\Tstrut
Rank & 99 & 100 & 99 & 99 & 99 & 99 & 99 & 99 & 99 & 99 \\
\hline
\Tstrut
FFT &  77 & 98 & 99 & 99 & 99 & 99 & 99 & 99 & 99 & 99\\
\hline 
\multicolumn{11}{l}{Non-overlappingTemplate} \Tstrut\\
\hspace{1cm}000000001 & \p 0 & \p 0 & \p 48 & 97 & 99 & 99 & 99 & 100 & 99 & 99 \\
\hspace{1cm}000000011 & \p 0 & \p 3 & 87 & 98 & 99 & 99 & 99 & 99 & 99 & 99 \\
\hspace{1cm}000000101 & \p 0 & \p 41 & 94 & 98 & 98 & 99 & 99 & 99 & 98 & 99 \\
\hspace{1cm}000000111 & \p 0 & \p 46 & 95 & 99 & 99 & 99 & 99 & 99 & 99 & 99 \\
\hspace{1cm}000001001 & \p 2 & 82 & 98 & 99 & 99 & 99 & 99 & 99 & 99 & 99 \\
\hspace{1cm}000001011 & \p 6 & 86 & 97 & 99 & 99 & 99 & 99 & 99 & 99 & 99 \\
\hspace{1cm}000001101 & \p 39 & 94 & 98 & 99 & 99 & 100 & 99 & 100 & 99 & 99 \\
\hspace{1cm}000001111 & \p 0 &74 & 97 & 99 & 99 & 99 & 99 & 99 & 99 & 99 \\
\hspace{1cm}000010001 &  68 & 95 & 99 & 98 & 99 & 99 & 99 & 100 & 99 & 99 \\
\hspace{1cm}000010011 & 76 & 95 & 98 & 99 & 99 & 99 & 99 & 99 & 98 & 99 \\
\hspace{1cm}000010101 & 94 & 97 & 99 & 99 & 99 & 99 & 99 & 99 & 99 & 99 \\
\hspace{1cm}000010111 & 64 & 95 & 98 & 99 & 99 & 99 & 99 & 99 & 98 & 99 \\
\hspace{1cm}000011001 & 96 & 98 & 99 & 99 & 100 & 100 & 99 & 99 & 98 & 99 \\
\hspace{1cm}000011011 & 94 & 98 & 99 & 99 & 99 & 99 & 99 & 99 & 99 & 99 \\
\hspace{1cm}000011101 & 96 & 98 & 98 & 99 & 99 & 99 & 99 & 99 & 99 & 99\\
\hline
\Tstrut
OverlappingTemplate & \p 0 & \p 0 & \p 11 &  93 & 98 & 99 & 98 & 99 & 99 & 99 \\
\hline
\Tstrut
Universal & \p 0 & 68 & 97 & 98 & 99 & 99 & 99 & 99 & 99 & 99  \\
ApproxEntropy $(m=10)$ & \p 0 & \p 0 & 64 & 98 & 99 & 99 & 99 & 100 & 99 & 99 \\
\hline
\multicolumn{11}{l}{RandomExcursions}\Tstrut \\
\hspace{1cm}$x= $ -4 & 90 & 98 & 99 & 99 & 98 & 99 & 99 & 99 & 99 & 100 \\
\hspace{1cm}$x=$ -3 & 91 & 97 & 99 & 99 & 99 & 99 & 99 & 99 & 99 & 99 \\
\hspace{1cm}$x=$ -2 & 94 & 99 & 98 & 99 & 99 & 98 & 99 & 99 & 99 & 99 \\
\hspace{1cm}$x=$ -1 & 95 & 99 & 99 & 98 & 99 & 99 & 99 & 99 & 99 & 100 \\
\hspace{1cm}$x=1$ &  95 & 99 & 99 & 99 & 99 & 99 & 99 & 98 & 99 & 100 \\
\hspace{1cm}$x=2$ &  93 & 98 & 99 & 99 & 98 & 100 & 98 & 98 & 99 & 99 \\
\hspace{1cm}$x=3$ &  89 & 97 & 98 & 99 & 99 & 99 & 98 & 98 & 99 & 99 \\
\hspace{1cm}$x=4$ &  85 & 97 & 97 & 98 & 98 & 99 & 99 & 98 & 98 & 100 \\

\hline
\multicolumn{11}{l}{RandExcursVar} \Tstrut\\
\hspace{1cm}$x=$ -9 & 99 & 100 & 99 & 99 & 100 & 99 & 99 & 99 & 99 & 100 \\
\hspace{1cm}$x=$ -8 & 99 & 99 & 99 & 99 & 99 & 99 & 99 & 99 & 99 & 100 \\
\hspace{1cm}$x=$ -7 & 100 & 99 & 99 & 99 & 99 & 99 & 99 & 99 & 99 & 100 \\
\hspace{1cm}$x=$ -6 & 100 & 99 & 99 & 99 & 99 & 99 & 98 & 99 & 99 & 99 \\
\hspace{1cm}$x=$ -5 & 100 & 99 & 99 & 99 & 99 & 99 & 99 & 99 & 99 & 99 \\
\hspace{1cm}$x=$ -4 & 99 & 100 & 99 & 99 & 99 & 99 & 99 & 99 & 99 & 99 \\
\hspace{1cm}$x=$ -3 & 99 & 100 & 99 & 98 & 99 & 99 & 99 & 99 & 99 & 99 \\
\hspace{1cm}$x=$ -2 & 99 & 99 & 99 & 98 & 99 & 99 & 99 & 100 & 99 & 99 \\
\hspace{1cm}$x=$ -1 & 99 & 99 & 99 & 99 & 99 & 99 & 99 & 99 & 99 & 99 \\
\hspace{1cm}$x=1$ & 100 & 100 & 99 & 100 & 100 & 99 & 99 & 99 & 99 & 99 \\
\hspace{1cm}$x=2$ & 99 & 99 & 99 & 100 & 99 & 99 & 99 & 99 & 99 & 100 \\
\hspace{1cm}$x=3$ & 99 & 99 & 98 & 99 & 99 & 99 & 99 & 99 & 99 & 100 \\
\hspace{1cm}$x=4$ & 99 & 99 & 99 & 100 & 99 & 99 & 99 & 99 & 99 & 99 \\
\hspace{1cm}$x=5$ & 99 & 99 & 98 & 100 & 98 & 99 & 98 & 99 & 99 & 99 \\
\hspace{1cm}$x=6$ & 99 & 98 & 98 & 99 & 98 & 99 & 99 & 99 & 99 & 99 \\
\hspace{1cm}$x=7$ & 99 & 98 & 99 & 100 & 98 & 100 & 99 & 99 & 99 & 99 \\
\hspace{1cm}$x=8$ & 99 & 99 & 100 & 99 & 99 & 99 & 99 & 100 & 99 & 99 \\
\hspace{1cm}$x=9$ & 99 & 98 & 100 & 99 & 98 & 99 & 99 & 100 & 99 & 99 \\
\hline
\multicolumn{11}{l}{Serial $(m=16)$} \Tstrut\\
\hspace{1cm} Serial 1 & \p 0 & \p1 & 82 & 96 & 98 & 99 & 99 & 98 & 99 & 99 \\
\hspace{1cm} Serial 2 & \p 10 &  81 &  95 & 98 & 98 & 99 & 99 & 99 & 98 & 99\\
\hline  \Tstrut
LinearComplexity $(M=500)$ & 99 & 98 & 99 & 99 & 99 & 99 & 99 & 98 & 99 & 99 \\
\bottomrule
\end{tabular}
\end{table}
The obtained number of files that passed the Diehard tests, i.e. $0.0001<\text{p-value}<0.9999$, are shown in Table~\ref{tab:diehardlogisticmap}. We observed that for some of the tests (see ranks6x8, monkey20bits, count1sStream and count1sSpecific) the logistic map from $k_0$ to $k_3$ failed badly (shown in gray). However, most of the 100 files passed successfully when $k\geq 4$. Regarding the failed tests, for instance, the monkey20bits test treats sequences of bits as ``words'', thus it counts the overlapping words in the stream looking for segments of 26-bits that do not appear from position 0 (leftmost) to 31. Therefore, as the logistic map from $k_0$ to $k_3$ maintains a probability distribution (U-shaped pattern), these bits segments are not encoded in the stream, which consequently influenced the test to fail.

\subsection{NIST battery}
\label{sec:nistlogistic}

The NIST test suite consists of fifteen statistical tests and the recommendations provided by the NIST [Special Publication 800-22 Revision 1a] were considered~\cite{NIST-PRNG}. A binary sequence is considered successful when the null hypothesis $p$-value $\geq 0.01$ is accepted. Then, 10 files containing 100Mbits were generated based on the $k$-logistic map orbit $\mathcal{O}^{k}(\mu,x_0)$, where $\mu\rightarrow 4$ as control parameter. Thus, 10 samples containing binary sequences of length $10^6$ bits were split from each file. There were 100   samples in total.

The number of samples that passed the NIST tests for each $k$-logistic map are reported in Table~\ref{tab:nistLogistic}. For the sake of simplicity, some of the sub-tests are not shown in the Table. This is similar to the results obtained from the DIEHARD tests. We observed a partial failure in the NIST test when using $k_0$, $k_1$, and $k_2$. More specifically, from top to down, i.e. runs, non-overlapping template, overlapping template, universal, approximate entropy and serial. Nevertheless, from $k\geq 3$, most of the 100 samples passed successfully.

\section{Extending to the k-tent map}

Although the focus of the article is mainly on the $k$-logistic map, our deep-zoom approach, which uses $k$-digits of precision can be extended to other chaotic maps in the unit interval such as the tent map, which is given by Eq.~\ref{eq:tentmap}.
%
\begin{equation}
  x_{t+1}= f(\delta,x_t)=
  \begin{cases}
   \delta x_t,& \text{if } x_t < 0.5\\
   \delta (1-x_t),& \text{if } x_t \geq 0.5,
	\end{cases}
\label{eq:tentmap}
\end{equation}
where $x_t\in M=[0,1]$ and $\delta \in U=[0,2]$. Therefore, we extended the definitions given in Section~\ref{Sec:proposal}. Thus, given an orbit of the original tent map $\mathcal{O}(\delta,x_0)= \{x_0,\ldots, x_t\}$ and the resulting orbit $\mathcal{O}^k(\delta,x_0)= \{x_0^k,\ldots, x_t^k\}$, we can generalize Eq.~\ref{eq:klogistic}.

In a similar way to the former experiments, here we also analyzed the overall dynamic properties of the $k$-tent map for purposes of the PRNGs. Thus, we also limited the experiments to the chaotic regime of the $k$-tent map by using the control parameter $\delta = 2$.

\subsection{Bifurcation diagram of the k-tent map}

\begin{figure}[!ht]
\centering
{\includegraphics[scale=0.276]{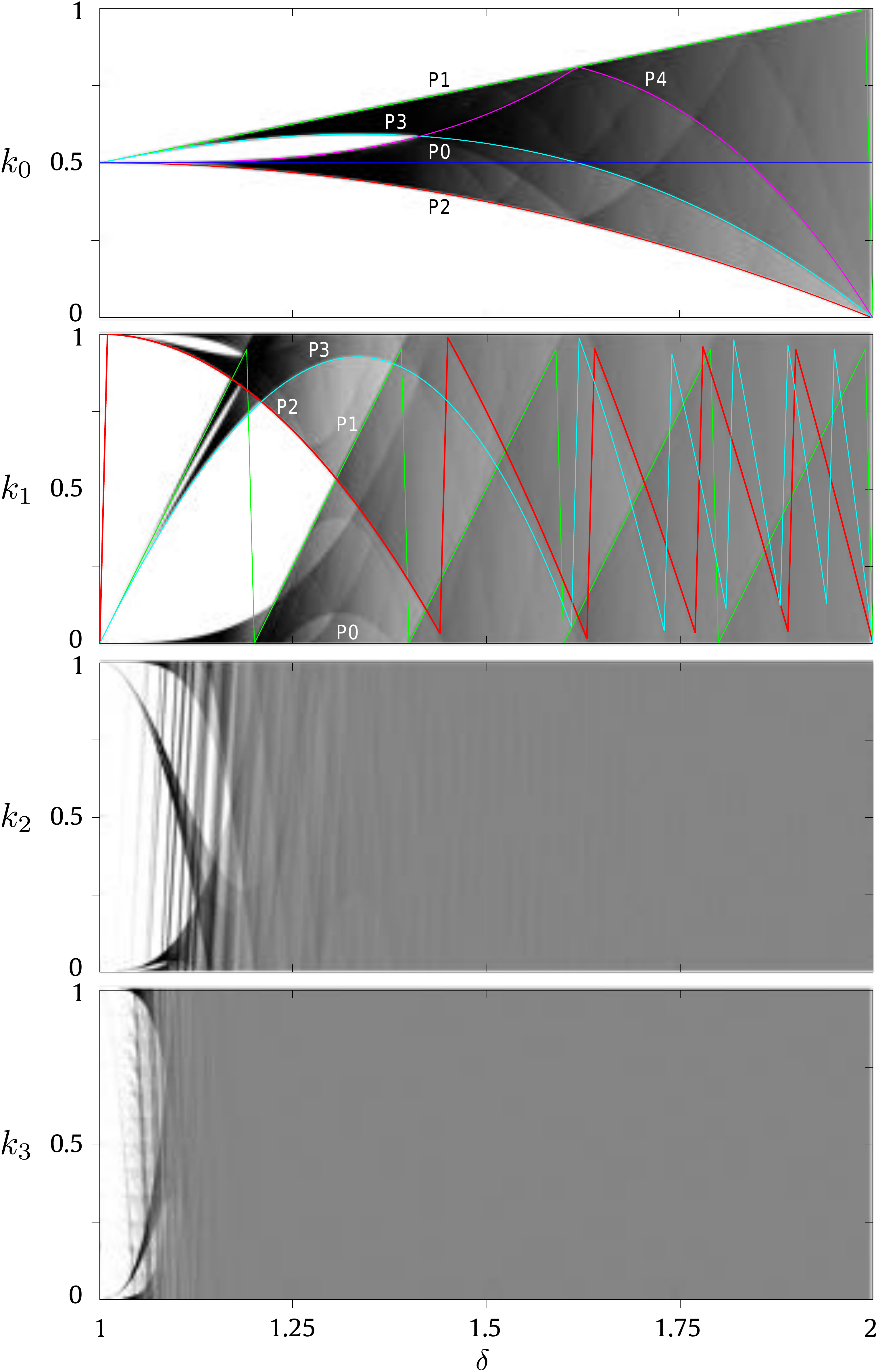}}
\caption{Bifurcation diagram of the $k_0$, $k_1$, $k_2$ and $k_3$-tent map (Eq.~\ref{eq:tentmap}). The horizontal axis shows $\mu\in [1,2]$ (steps of $0.001$), and the vertical axis depicts the long-term values of the $k$-tent map, starting from same initial condition. We calculated over $10^5$ iterations and discarded the first $200$ transient values. The superimposed curves $P_{n+1}(\delta)$ correspond to the skeleton bifurcation (dark lines) regarding the tent map, where $P_{n+1}(\delta)= f(P_{n}, \delta)$ with $x_c= 0.5$.}
\label{fig:biffkmapatent}
\end{figure}

Moreover, Fig.~\ref{fig:biffkmapatent} shows the bifurcation and skeleton diagram for the $k$-tent map. We observed a similar phenomenon as reported in the $k$-logistic map (Fig.~\ref{fig:biffkmapalogistico}). Therefore, as the $k$ digits are increased, the orbits become twisted rapidly, the forks of periodicity shrink and the steady regions remain. Moreover, the superimposed skeleton bifurcation also shows a zigzag pattern behavior which refers to a more irregular pattern.

\subsection{Phase diagram of the k-tent map}

\begin{figure}[!ht]
  \centering
{\includegraphics[width=7.5cm,height=15cm]{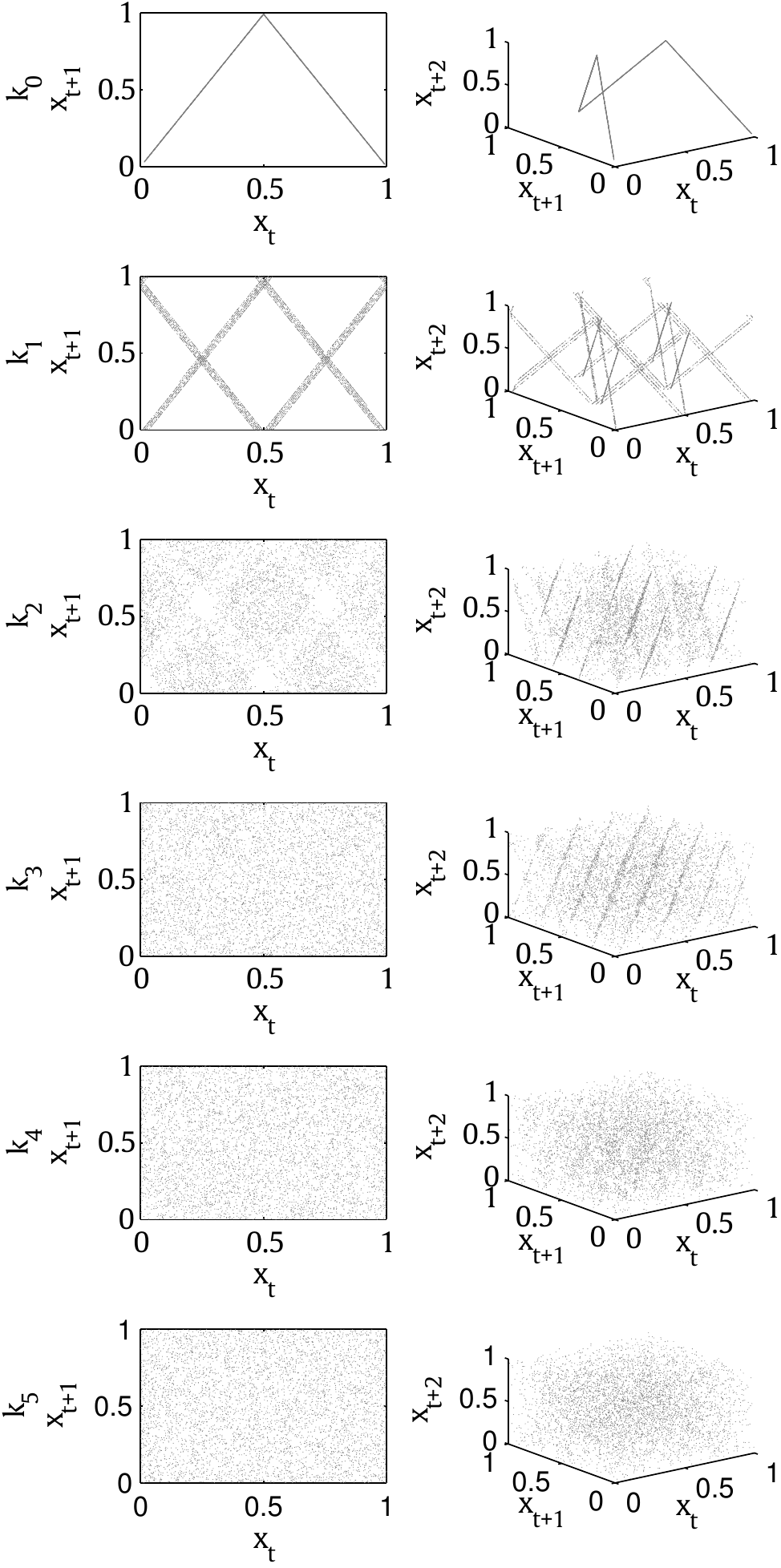}}
  \caption{Phase diagrams for the $k_0$, $k_1$, $k_2$, $k_3$, $k_4$ and $k_5$ using $\delta = 2$. Two- and three-dimensional phase diagrams are shown on the left and right, respectively. The horizontal and vertical axes show the phase space of $x_t$ against $x_{t+1}$ and $x_{t+2}$. Each orbit contains $10^4$ points from a random initial condition. The first 200 were discarded (transient period).}
  \label{fig:poincareTent}
\end{figure}

The phase diagrams of the $k$-tent map are shown in Fig.~\ref{fig:poincareTent} (from top to down) corresponding to the $k_0$ to $k_5$. We noticed that the $k$-tent map behaves similarly to the logistic map (see Fig.~\ref{fig:poincareLogistic}) in terms of the randomization course as $k$ is increased. Here, the characteristic tent pattern ($k_0$), from which it received its name, seems to zigzag ($k_1$), until it become more diffused. Nevertheless, some fuzzy patterns still remain at the three-dimensional phase diagram of $k_4$ and the phase-space begins to randomize from $k\geq 5$.

\subsection{PRNG based on k-tent map}

We also used the $k$-tent map for PRNG purposes. Similar to the $k$-logistic map, here we restricted the phase-space $M$ to the interval $]0,1[$ and the control parameter to the chaotic windows, e.g $\delta \rightarrow 2$. Thus, in order to generate numbers, we discretize the $x_t^k$ values from the $k$-tent map into 32-bit integers as defined in Eq.~\ref{eq:discretize}. We set up the same configuration for the experiments using the Diehard~\cite{DIEHARD} and NIST~\cite{NIST-PRNG} suite tests. Table~\ref{tab:diehardtentmap} and Table~\ref{tab:NIST_TENT} reported the number of files that have passed the pseudo-randomness tests to both suites, respectively.

Regarding these tables, we observed a partial failure in the Diehard test from the $k_0$ to $k_5$-tent map. More precisely, the sub-tests Ranks6x8 matrices, OQSO, Count1sStream and Count 1s specifics. Notwithstanding from $k\geq 6$, most of the 100 samples passed successfully. In contrast to the $k$-logistic map, the $k$-tent map successfully passed the NIST test for all of the $k$ parameters.

\begin{table}[!ht]
\label{tab:diehardtentmap}
\caption{Average number of files that passed testes DIEHARD using $k$-tent map PRNG (Eq.~\ref{eq:discretize}) from 100 files samples. Severe failed tests are shown in gray. All tests are passed at the interval $0.0001<\text{p-value}<0.9999$.}
\small
\centering
\begin{tabular}{lrrrrrrrrrr}
\toprule
\textbf{Diehard tests} & \multicolumn{1}{c}{$k_0$} & \multicolumn{1}{c}{$k_1$} & \multicolumn{1}{c}{$k_2$} & \multicolumn{1}{c}{$k_3$} & \multicolumn{1}{c}{$k_4$} & \multicolumn{1}{c}{$k_5$} & \multicolumn{1}{c}{$k_6$} & \multicolumn{1}{c}{$k_7$} & \multicolumn{1}{c}{$k_8$} & \multicolumn{1}{c}{$k_9$} \Tstrut\\
\hline \Tstrut
BirthdaySpacings [KS] & 100 & 100 & 100 & 100 & 100 & 100 & 100 & 99 & 100 & 100 \\
OverlappingPermutations & 99 & 97 & 100 & 99 & 96 & 98 & 97 & 98 & 100 & 97 \\
Ranks31x31 matrices & 100 & 100 & 100 & 100 & 100 & 100 & 100 & 100 & 100 & 100 \\
Ranks32x32 matrices & 100 & 100 & 100 & 100 & 100 & 100 & 100 & 100 & 100 & 100 \\
Ranks6x8 matrices [KS] & \pp0 & \pp0 & \pp0 & 100 & 100 & 100 & 100 & 100 & 100 & 100 \\
Monkey20bitsWords [KS] & 100 & 100 & 100 & 100 & 100 & 100 & 100 & 100 & 100 & 100 \\
OPSO [KS] & 100 & 100 & 100 & 100 & 100 & 100 & 100 & 100 & 100 & 100 \\
OQSO [KS] & \pp0 & \pp0 & \pp0 & 100 & 100 & 100 & 100 & 100 & 100 & 100 \\
DNA [KS] & 100 & 100 & 100 & 100 & 100 & 100 & 100 & 100 & 100 & 100 \\
Count1sStream & \pp0 & \pp0 & \pp0 & \pp0 & 100 & 100 & 100 & 100 & 100 & 100 \\
Count1sSpecific [KS] & \pp0 & \pp0 & \pp0 & \pp0 & \pp0 & \pp0 & 100 & 100 & 100 & 100 \\
ParkingLot [KS] & 100 & 100 & 100 & 100 & 100 & 100 & 100 & 100 & 100 & 100 \\
MinimumDistance [KS] & 81 & 95 & 94 & 100 & 98 & 98 & 97 & 99 & 98 & 99 \\
RandomSpheres [KS] & 100 & 100 & 100 & 100 & 100 & 100 & 100 & 100 & 100 & 100 \\
Squeeze [KS] & 100 & 100 & 100 & 100 & 100 & 100 & 100 & 100 & 100 & 100 \\
OverlappingSums [KS] & 100 & 100 & 100 & 100 & 100 & 100 & 100 & 100 & 100 & 100 \\
Runs (up) & 100 & 100 & 100 & 100 & 100 & 100 & 100 & 100 & 100 & 100 \\
Runs (down) & 100 & 100 & 100 & 100 & 100 & 100 & 100 & 100 & 100 & 100 \\
Craps (wins) & 100 & 100 & 100 & 100 & 100 & 100 & 100 & 100 & 100 & 100 \\
Craps (throws/game) & 100 & 100 & 100 & 100 & 100 & 100 & 100 & 100 & 100 & 100\\
\bottomrule
\end{tabular}
\end{table}

\setlength{\tabcolsep}{2pt}
\begin{table}[!ht]
\caption{Number of files that passed the NIST test suits~\cite{NIST-PRNG} for $k$-tent map.  All tests are passed at the $\alpha=0.01$ significance level.}
\label{tab:NIST_TENT}
\fontsize{9}{9}\selectfont
\centering


\begin{tabular}{lrrrrrrrrrr}
\toprule
NIST tests &$k_0$ &$k_1$ &$k_2$ &$k_3$ &$k_4$ &$k_5$ &$k_6$ &$k_7$ &$k_8$ &$k_9$ \\
\hline
\Tstrut
Frequency  & 98  & 99  & 99  & 99  & 99  & 99  & 98  & 99  & 99  & 99  \\
\Tstrut
BlockFrequency $(m=128)$ & 100 & 99 & 99 & 100 & 99 & 100 & 99 & 100 & 99 & 99 \Tstrut\\
\hline  
\multicolumn{11}{l}{CumulativeSums}  \Tstrut\\
\hspace{1cm}Forward sums & 98 & 99 & 99 & 98 & 99 & 99 & 98 & 99 & 99 & 99 \\
\hspace{1cm}Reverse sums & 98 & 99 & 99 & 99 & 99 & 99 & 99 & 99 & 99 & 99 \\
\hline 
\Tstrut
Runs &  79 & 92 & 98 & 99 & 99 & 99 & 99 & 99 & 99 & 99 \\
\Tstrut
LongestRun & 97 & 99 & 99 & 99 & 99 & 99 & 99 & 99 & 99 & 99 \\
\hline
\Tstrut
Rank & 99 & 99 & 99 & 99 & 99 & 99 & 99 & 99 & 100 & 99 \\
\hline
\Tstrut
FFT & 99 & 99 & 99 & 99 & 99 & 99 & 99 & 99 & 98 & 99 \\
\hline 
\multicolumn{11}{l}{Non-overlappingTemplate} \Tstrut\\
\hspace{1cm}000000001 & 75 & 97 & 99 & 99 & 99 & 99 & 99 & 99 & 98 & 99 \\
\hspace{1cm}000000011 & 89 & 97 & 99 & 100 & 99 & 99 & 98 & 99 & 99 & 99 \\
\hspace{1cm}000000101 & 92 & 98 & 99 & 99 & 100 & 99 & 100 & 98 & 99 & 99 \\
\hspace{1cm}000000111 & 91 & 99 & 99 & 99 & 99 & 99 & 99 & 99 & 99 & 99 \\
\hspace{1cm}000001001 & 92 & 99 & 99 & 99 & 99 & 99 & 100 & 99 & 99 & 99 \\
\hspace{1cm}000001011 & 93 & 99 & 99 & 99 & 99 & 99 & 99 & 99 & 99 & 99 \\
\hspace{1cm}000001101 & 95 & 99 & 99 & 99 & 99 & 100 & 99 & 99 & 99 & 99 \\
\hspace{1cm}000001111 & 90 & 98 & 99 & 99 & 99 & 99 & 99 & 99 & 99 & 99 \\
\hspace{1cm}000010001 & 93 & 100 & 99 & 99 & 99 & 98 & 99 & 99 & 99 & 99 \\
\hspace{1cm}000010011 & 93 & 99 & 98 & 99 & 99 & 99 & 99 & 99 & 99 & 99 \\
\hspace{1cm}000010101 & 94 & 99 & 99 & 99 & 99 & 99 & 99 & 99 & 99 & 100 \\
\hspace{1cm}000010111 & 93 & 99 & 99 & 99 & 99 & 99 & 99 & 99 & 99 & 99 \\
\hspace{1cm}000011001 & 93 & 99 & 99 & 99 & 99 & 99 & 100 & 100 & 99 & 99 \\
\hspace{1cm}000011011 & 94 & 98 & 99 & 99 & 99 & 99 & 99 & 99 & 99 & 99 \\
\hspace{1cm}000011101 & 93 & 98 & 99 & 99 & 99 & 99 & 99 & 99 & 99 & 99 \\
\hline
\Tstrut
OverlappingTemplate & 62 & 91 & 99 & 99 & 99 & 99 & 99 & 99 & 99 & 99  \\
\hline
\Tstrut
Universal & 97 & 98 & 99 & 99 & 99 & 99 & 99 & 99 & 98 & 99   \\
ApproxEntropy $(m=10)$ & 82 & 99 & 99 & 99 & 99 & 99 & 99 & 99 & 99 & 99  \\
\hline
\multicolumn{11}{l}{RandomExcursions}\Tstrut \\
\hspace{1cm}$x=$-4 & 98 & 99 & 99 & 98 & 99 & 100 & 98 & 98 & 99 & 99 \\
\hspace{1cm}$x=$-3 & 100 & 98 & 99 & 98 & 99 & 99 & 99 & 99 & 99 & 99 \\
\hspace{1cm}$x=$-2 & 98 & 99 & 99 & 99 & 99 & 99 & 99 & 99 & 100 & 99 \\
\hspace{1cm}$x=$-1 & 99 & 99 & 99 & 99 & 98 & 99 & 99 & 99 & 99 & 99 \\
\hspace{1cm}$x=1$ & 99 & 99 & 99 & 99 & 99 & 98 & 100 & 99 & 99 & 99 \\
\hspace{1cm}$x=2$ & 100 & 98 & 99 & 99 & 99 & 100 & 99 & 99 & 98 & 99 \\
\hspace{1cm}$x=3$ & 99 & 99 & 99 & 99 & 99 & 100 & 99 & 99 & 99 & 99 \\
\hspace{1cm}$x=4$ & 98 & 99 & 98 & 99 & 99 & 99 & 99 & 99 & 99 & 99 \\

\hline
\multicolumn{11}{l}{RandExcursVar} \Tstrut\\
\hspace{1cm}$x=$-9 & 99 & 100 & 99 & 99 & 100 & 99 & 99 & 99 & 99 & 98 \\
\hspace{1cm}$x=$-8 & 99 & 99 & 99 & 100 & 100 & 99 & 100 & 99 & 99 & 98 \\
\hspace{1cm}$x=$-7 & 99 & 99 & 99 & 100 & 99 & 99 & 99 & 100 & 99 & 98 \\
\hspace{1cm}$x=$-6 & 99 & 99 & 100 & 100 & 100 & 99 & 99 & 99 & 98 & 99 \\
\hspace{1cm}$x=$-5 & 99 & 99 & 99 & 100 & 100 & 99 & 99 & 99 & 99 & 98 \\
\hspace{1cm}$x=$-4 & 99 & 99 & 99 & 99 & 99 & 99 & 99 & 99 & 100 & 99 \\
\hspace{1cm}$x=$-3 & 99 & 99 & 100 & 99 & 100 & 99 & 99 & 100 & 99 & 99 \\
\hspace{1cm}$x=$-2 & 99 & 99 & 100 & 100 & 99 & 99 & 100 & 99 & 99 & 99 \\
\hspace{1cm}$x=$-1 & 99 & 99 & 99 & 99 & 99 & 99 & 99 & 99 & 98 & 99 \\
\hspace{1cm}$x=1$ & 99 & 99 & 98 & 99 & 99 & 99 & 98 & 100 & 98 & 99 \\
\hspace{1cm}$x=2$ & 98 & 99 & 98 & 99 & 99 & 99 & 98 & 99 & 98 & 99 \\
\hspace{1cm}$x=3$ & 98 & 99 & 99 & 99 & 99 & 100 & 98 & 99 & 99 & 99 \\
\hspace{1cm}$x=4$ & 99 & 99 & 98 & 99 & 98 & 99 & 99 & 99 & 99 & 99 \\
\hspace{1cm}$x=5$ & 98 & 99 & 98 & 99 & 98 & 99 & 98 & 99 & 99 & 99 \\
\hspace{1cm}$x=6$ & 98 & 99 & 99 & 99 & 99 & 99 & 99 & 99 & 99 & 99 \\
\hspace{1cm}$x=7$ & 99 & 100 & 100 & 99 & 99 & 99 & 98 & 99 & 99 & 99 \\
\hspace{1cm}$x=8$ & 99 & 99 & 99 & 99 & 99 & 99 & 99 & 99 & 99 & 99 \\
\hspace{1cm}$x=9$ & 99 & 99 & 99 & 99 & 99 & 99 & 98 & 99 & 99 & 99 \\
\hline
\multicolumn{11}{l}{Serial $(m=16)$} \Tstrut\\
\hspace{1cm} Serial 1 & 99 & 99 & 99 & 99 & 99 & 99 & 99 & 99 & 99 & 99 \\
\hspace{1cm} Serial 2 & 99 & 99 & 99 & 99 & 100 & 99 & 99 & 99 & 99 & 99 \\
\hline  \Tstrut
LinearComplexity $(M=500)$ & 99 & 99 & 99 & 99 & 99 & 100 & 99 & 99 & 98 & 99 \\
\bottomrule
\end{tabular}
\end{table}

\subsection{Comparison with other PRNGs}

In order to compare the performance of the pseudo-randomness properties of the $k$-logistic map and the $k$-tent map with classical PRNGs, we chose the most popular PRNGs, such as the congruential linear generator (LCG)~\cite{KnuthBook,PressPRNG} and the Mersenne Twister (MT)~\cite{MersenneTwister}. The LCG is defined by the recurrence relation given by $x_{n+1}=\left(ax_{n}+c\right)~~{\bmod {~}}~m$, where the parameters $a$, $c$ and $m$ correspond to the integer parameters (the added, the constant, the multiplier) of the LCG generator. Here, we implemented the Java LCG version, using $a=11$, $c=25214903917$, $m=2^{48}$. Furthermore, we used the Java version\footnote{Copyright (c) 2003 by Sean Luke \url{https://cs.gmu.edu/~sean/research/mersenne}} of the original Mersenne Twister PRNG~\cite{MersenneTwister}.

The average number of files that passed to the Diehard test is summarized in Table~\ref{tab:comparediehard}. This Table shows comparable results obtained for the LCG and the MT PRNGs with the $k_4$-logistic map.

\begin{table}[!ht]
\label{tab:comparediehard}
\caption{Comparison of the avg. number of passed testes DIEHARD using the $k_4$-logistic map PRNG and the congruential linear generator (LCG) and the Mersenne Twister PRNG (MT). Severe failed tests are shown in gray. All tests are passed at the interval $0.0001<\text{p-value}<0.9999$.}
\small
\centering
\begin{tabular}{lrrr}
\toprule
\textbf{Diehard tests}  & logistic map ($k_4$) & LCG  &MT  \Tstrut\\
\hline \Tstrut
BirthdaySpacings [KS] & 100 & 100 & 100 \\
OverlappingPermutations & 98 & 99 & 94 \\
Ranks31x31 matrices & 100 & 100 & 100 \\
Ranks32x32 matrices & 100 & 100 & 100 \\
Ranks6x8 matrices [KS] & 100 & 100 & 100 \\
Monkey20bitsWords [KS] & 100 & 100 & 100 \\
OPSO [KS] & 100 & 58 & 100 \\
OQSO [KS] & 100 & \pp0 & 100 \\
DNA [KS] & 100 & \pp0 & 100 \\
Count1sStream & 100 & 100 & 100 \\
Count1sSpecific [KS] & 94 & 100 & 100 \\
ParkingLot [KS] & 100 & 100 & 100 \\
MinimumDistance [KS] & 100 & 100 & 100 \\
RandomSpheres [KS] & 100 & 100 & 100 \\
Squeeze [KS] & 100 & 100 & 100 \\
OverlappingSums [KS] & 100 & 100 & 100 \\
Runs (up) & 100 & 100 & 100 \\
Runs (down) & 100 & 100 & 100 \\
Craps (wins) & 100 & 100 & 100 \\
Craps (throws/game) & 100 & 100 & 100\\
\bottomrule
\end{tabular}
\end{table}

\begin{figure}[!ht]
\centering
{\includegraphics[scale=0.6]{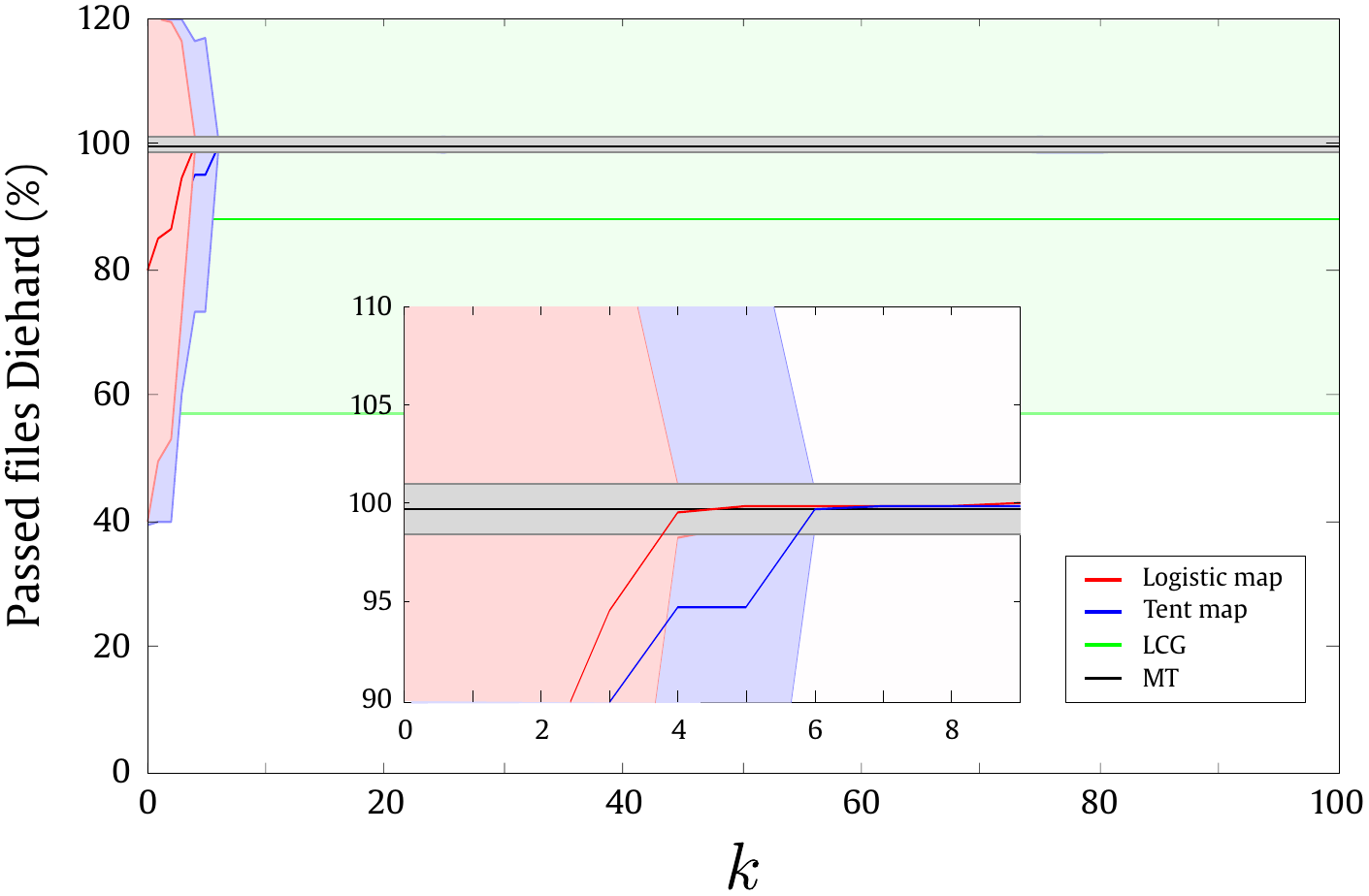}}
\caption{Average number of files that passed the Diehard test of the $k$-logistic map, $k$-tent map for $k_0$ to $k_{100}$, linear congruential generator (LCG) and Mersenne Twister (MT) PRNGs. The vertical axis indicates the average number while the horizontal axis the different values of $k$-digits. The curves represent the mean and standard deviation (shaded error bar). The inner plot shows the zoom of the region between $k \in [0, 9]$.}
\label{fig:diehardcomparison}
\end{figure}

Moreover, Fig.~\ref{fig:diehardcomparison} depicts the comparison of the average number of passed files from the Diehard suit of the $k$-logistic map, $k$-tent map from $k_0$ to $k_{100}$, LCG and MT generators. Therefore, based on the Student's t-test, we compare the means of all of the results from the Diehard test with a confidence level of 95\%. We observe that the logistic map outperformed the tent map regarding the battery of pseudo-randomness tests. Moreover, concerning the inner plot in Fig.~\ref{fig:diehardcomparison}, when comparing our approach to the LCG and the MT PRNGs, we observed that from $k\geq 3$ the logistic map obtained more pseudo-randomness properties than the LCG, while for $k\geq 3$ the logistic map performed just as well as the Mersenne Twister PRNG. 

\section{Discussions}
\label{sec:discussions}

In this paper, we focused on the logistic map to analyze the properties that arise from the $k$-right digit zoom. It was selected due to its simplicity and the fact that it is a well-known chaotic system. Therefore, we conducted two types of analysis: (i) to investigate the overall dynamic properties of the map and (ii) to assess the pseudo-randomness quality of the proposed PRNG by using both statistical tests, i.e. Diehard and NIST suites. 

In the first part, we analyzed the time-evolution (Fig.~\ref{fig:timeevolution}), the bifurcation diagram (Fig.~\ref{fig:biffkmapalogistico}), the Lyapunov exponent (Fig.~\ref{fig:biffkmapaLE}), the Poincar{\'e} diagram (Fig.~\ref{fig:poincareLogistic}) and the frequency distribution (Fig.~\ref{fig:histogramas}). 

The time-evolution analysis demonstrated that the $k$-logistic map keeps the overall dynamics of the original logistic map. Therefore, the chaotic, periodic and stable regions are preserved, regardless of parameter $k$. Indeed, what was changed was the number of iterations to stabilize the system. Thus, within the chaotic regions, the number of iterations before they diverge decreases with $k$, while within the stable regions, the number of iterations that converge to a fixed point and the cycle length of the periodic regions increases with $k$.

Furthermore, the analysis of the bifurcation diagram and the Poincar{\'e} diagram reveal a fast randomization behavior as $k$ increases. We observed an intensification of the chaotic behavior, which was corroborated by the Lyapunov exponent as all the chaotic regions tended to achieve the maximal LE upper bound. Regarding the frequency distribution analysis, we also observed a rapid randomization as $k$ increases, since the U pattern achieved a plateau distribution, which in principle indicates that the $k$-orbits visit virtually every region of the phase-space. In all of these analyses, we found that the pseudo-random properties of the $k$-logistic map can be noticeably improved as $k$ is increased. In fact, the patterns become more and more diffused until becoming visually indistinguishable ($k\geq4$).

In the second part of the experiments, regarding the pseudo-randomness tests, we found that the sequences generated when using the $k\geq 4$-logistic map and $k\geq 5$-tent map passed successfully to both the Diehard and NIST randomness tests. 

\section{Conclusions}
\label{sec:conclusions}

In this paper, we proposed an approach to increase the pseudo-randomness properties of a chaos-based PRNG by means of a ``deep-zoom'' exploration. The main idea is to compound new values from the $k$-right digits, after the decimal separator, from the outcomes of the map, to then expand them to a larger number of pseudo random bits. The deep-zoom exploration was motivated by the fact that the chaotic systems are analytically based on the infinitesimal digits of precision, of which presumably an infinite number of patterns can be extracted.

We proposed a general approach that can be directly applied to any one-dimensional chaotic map in the unit interval (i.e $x_t \in [0,1]$). Nevertheless, it can be adapted to other chaotic systems that do not satisfy the former condition, such as the H{\'e}non map, Gaussian map, Lorenz attractor, etc, by properly adjusting the scale of the phase-space, which implies composing new values $x_t^k$ in the same interval of the phase-space $M\in \mathbb{R}$.

Throughout this manuscript, we found that the pseudo-randomness properties of a chaotic map can be improved as $k$ is increased, i.e. a rapid transition from a ``weak'' to a ``strong'' randomness within the $k$-right sequences. Indeed, the overall dynamic analysis and the pseudo-randomness tests suggest that the quality performance of the proposed PRNG settle down in $k\geq4$ and $k\geq 5$ for the logistic map and tent map, respectively.

Furthermore, we also observed that our proposed nonlinear PRNG can be compared to the popular Mersenne Twister PRNG in terms of the pseudo-randomness quality provided by the statistical tests Diehard and NIST although these qualifications should not be considered as an ultimate indicator, but rather as a detector of deviations from randomness. 
This is due to the fact that these tests are supported statistically  considering the pass and fail values  instead of numerical values  that may lead to a measure of quality from the weak to the strong.
Therefore, there is still a lack of methods that show a measurement of a PRNG quality in terms of pseudo-randomness tests.

Moreover,  the PRNG requirements were reached, as uniform distribution,  long periodicity and indistinguishable statistical patterns were achieved. In fact, the proposed PRNG based on the deep-zoom of chaotic maps can be a good candidate for potential applications such as simulations, computer games, programming language, cryptography, among others.

Finally, in this manuscript we showed that our approach applied to a simple chaotic map is able to produce pseudo-random numbers of high-quality by taking advantage of their deep-zoom properties.

\section*{Acknowledgments}
J.M. acknowledges a scholarship from the Brazilian agency CAPES (PROEX). O.M.B. acknowledges support from CNPq (Grants \#307797/2014-7 and \#484312/2013-8) and FAPESP (Grant \#14/08026-1).

\appendix
\section*{Appendix A: Skeleton bifurcation}
\label{sec:AppendixA}

Bai-lin~\cite{HaoBailinBook} introduced a family of recursive functions $\{P_n(\mu)\}$, where $n=0,1, \ldots$, which correspond to the skeleton curves from a discrete dynamical system shown in a bifurcation diagram. The skeleton curves correspond to the characteristic dark lines and boundary zones as seen in the bifurcation diagrams shown in the logistic map (Fig.~\ref{fig:biffkmapalogistico}) and the tent map (Fig.~\ref{fig:biffkmapatent}). Here, we have shown the means to achieve the equation given at Eq.~\ref{eq:skeleton} as follows: 

\begin{definition}
 Considering that a DDS could have one or more critical points $x_{c_i}$, i.e. where its first derivative vanishes, i.e., $f^\prime(\mu,x_{c_i})=0$, many composite functions $\{P_n^i\}$ can be obtained, where the superscript $i$ indicates the related point. Thus, starting from one critical point $x_c$, it is defined recursively in the form $P_{n+1}= f(\mu, P_n(\mu))$. Regarding the logistic map, given its unique critical point $x_c=0.5$, since its first derivative is $f^\prime(\mu,x_c)=\mu -2\mu x = 0$, then the function $P_n$ are polynomials of $\mu$ such that:

\begin{equation*}
\begin{aligned}
P_0(\mu) &= x_c \\
P_1(\mu) &= \mu/4 \\
P_2(\mu) &= 4\mu^2-\mu^3/16 \\
P_3(\mu) &= (\mu^3- \mu^4)(\mu^3 - 4\mu^2 + 16)/256\\
P_4(\mu) &= (4\mu^4-\mu^5)(\mu^3 - 4\mu^2 + 16)\\
&(\mu^7 - 8\mu^6 + 16\mu^5 + 16\mu^4 - 64\mu^3 + 256))/65536\\
\end{aligned}
\end{equation*}

Furthermore, regarding the tent map and its unique critical point $x_c=0.5$, the polynomials $P_n(\delta)$ are as follows:

\begin{equation*}
\left.\begin{aligned}
P_0(\delta) &= x_c \\
P_1(\delta) &= \delta/2 \\
P_2(\delta) &= \left \{
 \begin{aligned}
  &\delta^2/2, & \text{ for }\ \delta < 1 \\
  &\delta(1- \delta)/2, & \text{ for }\ \delta \geq 1
 \end{aligned} \right. \\
\end{aligned}\right.
\end{equation*}

\end{definition}


\end{document}